\pdfoutput=1

\documentclass[aps,prl,longbibliography,reprint,amssymb,amsmath,floatfix,superscriptaddress]{revtex4-1}

\usepackage[utf8]{inputenc}
\usepackage[T1]{fontenc}

\usepackage{newtxtext}
\usepackage[varvw]{newtxmath}
\usepackage{textcomp}
\usepackage{newtxmath}
\usepackage{bm}
\usepackage{amsmath}

\usepackage[english]{babel}
\usepackage{csquotes}

\usepackage{microtype}

\usepackage{graphicx}

\usepackage{float}
\usepackage{flafter}
\usepackage{pgf}
\usepackage{tikz}

\usepackage{siunitx}
\usepackage[version=3]{mhchem}

\usepackage{xcolor}
\definecolor{DarkRed}{rgb}{0.65,0,0}
\definecolor{DarkBlue}{rgb}{0,0,0.65}
\definecolor{DarkPurple}{rgb}{0.65,0,0.65}
\definecolor{DarkGreen}{rgb}{0,0.65,0}
\definecolor{DarkYellow}{rgb}{0,0.65,0.65}

\usepackage[colorlinks=true,allcolors=blue]{hyperref}
\addto\extrasenglish{\def\equationautorefname~#1\null{Eq.~(#1)\null}}
\AtBeginDocument{

}

\newcommand{\ve}[1]{\bm{#1}}

\newcommand{\diff}[1]{\mathrm{d} #1}
\DeclareMathOperator{\real}{Re}

\providecommand{\abs}[1]{\lvert#1\rvert} 

\providecommand{\ie}{\textit{i.e. }}
\providecommand{\eg}{\textit{e.g. }}
\providecommand{\prlsection}[1]{\textit{#1}.\kern.05em---\kern.05em\ignorespaces}

\begin{document}

\title{Magnon-Mediated Indirect Exciton Condensation through Antiferromagnetic Insulators}
\author{Øyvind Johansen}
\email{oyvinjoh@ntnu.no}
\affiliation{Center for Quantum Spintronics, Department of Physics, Norwegian University of Science and Technology, NO-7491 Trondheim, Norway}
\author{Akashdeep Kamra}
\affiliation{Center for Quantum Spintronics, Department of Physics, Norwegian University of Science and Technology, NO-7491 Trondheim, Norway}
\author{Camilo Ulloa}
\affiliation{Institute for Theoretical Physics, Utrecht University, Princetonplein 5,
3584 CC Utrecht, the Netherlands}
\author{Arne Brataas}
\affiliation{Center for Quantum Spintronics, Department of Physics, Norwegian University of Science and Technology, NO-7491 Trondheim, Norway}
\author{Rembert A. Duine}
\affiliation{Center for Quantum Spintronics, Department of Physics, Norwegian University of Science and Technology, NO-7491 Trondheim, Norway}
\affiliation{Institute for Theoretical Physics, Utrecht University, Princetonplein 5,
3584 CC Utrecht, the Netherlands}
\affiliation{Department of Applied Physics, Eindhoven University of Technology, P.O. Box 513, 5600 MB Eindhoven, The Netherlands}
\date{\today}

\begin{abstract}
Electrons and holes residing on the opposing sides of an insulating barrier and experiencing an attractive Coulomb interaction can spontaneously form a coherent state known as an indirect exciton condensate. We study a trilayer system where the barrier is an antiferromagnetic insulator. The electrons and holes here additionally interact via interfacial coupling to the antiferromagnetic magnons. We show that by employing magnetically uncompensated interfaces, we can design the magnon-mediated interaction to be attractive or repulsive by varying the thickness of the antiferromagnetic insulator by a single atomic layer. We derive an analytical expression for the critical temperature $T_c$ of the indirect exciton condensation. Within our model, anisotropy is found to be crucial for achieving a finite $T_c$, which increases with the strength of the exchange interaction in the antiferromagnetic bulk. For realistic material parameters, we estimate $T_c$ to be around \SI{7}{\kelvin}, the same order of magnitude as the current experimentally achievable exciton condensation where the attraction is solely due to the Coulomb interaction. The magnon-mediated interaction is expected to cooperate with the Coulomb interaction for condensation of indirect excitons, thereby providing a means to significantly increase the exciton condensation temperature range.
\end{abstract}

\maketitle

\prlsection{Introduction}
Interactions between fermions result in exotic states of matter. Superconductivity is a prime example, where the negatively charged electrons can have an overall attractive coupling mediated by individual couplings to the vibrations, known as phonons, of the positively charged lattice. In addition to charge, the electron also has a spin degree of freedom.
The electron spin can interact with localized magnetic moments through an exchange interaction exciting the magnetic moment by transfer of angular momentum. These excitations are quasiparticles known as magnons.
Theoretical predictions of electron-magnon interactions have shown that these can also induce effects such as superconductivity \cite{Suhl:prl:2001,Karchev:prb:2003,Funaki2014,Kar2014,Kargarian:prl:2016,Gong2017,Rohling:prb:2018,Hugdal:prb:2018,Erlandsen2019,Fjaerbu2019}.

Research interest in antiferromagnetic materials is surging \cite{Jungwirth2016,Baltz2018}. This enthusiasm is due to the promising properties of antiferromagnets such as high resonance frequencies in the THz regime and a vanishing net magnetic moment. Much of this research focuses on interactions involving magnons or spin waves at magnetic interfaces in hybrid structures. Examples of this are spin pumping \cite{Tserkovniak:prl:2002,Ross:TUMunchen:2013,Cheng:prl:2014,Takei:prb:2014,Ross:2015,Johansen:prb:2017,Kamra:prl:2017}, spin transfer \cite{Cheng:prl:2014,Cheng:prl:2016,Sluka:prb:2017,Johansen:prb:2018}, and spin Hall magnetoresistance \cite{Han:prb:2014,Hou:prl:2017,Hoogeboom:apl:2017,Manchon2017,Fischer:prb:2018,Baldrati:prb:2018} at normal metal interfaces, and magnon-mediated superconductivity \cite{Erlandsen2019,Fjaerbu2019}. Recently, an experiment has also demonstrated spin transport in an antiferromagnetic insulator over distances up to \SI{80}{\micro\metre} \cite{Lebrun2018}. Moreover, antiferromagnetic materials are also of interest since it is believed that high-temperature superconductivity in cuprates is intricately linked to magnetic fluctuations near an antiferromagnetic Mott insulating phase \cite{Bonn2006,HighTc}. Thus it is crucial to achieve a good understanding of antiferromagnetic magnon-electron interactions, as well as electron-electron interactions mediated by antiferromagnetic magnons.

\begin{figure}[b]
\centering
\includegraphics[width=\linewidth]{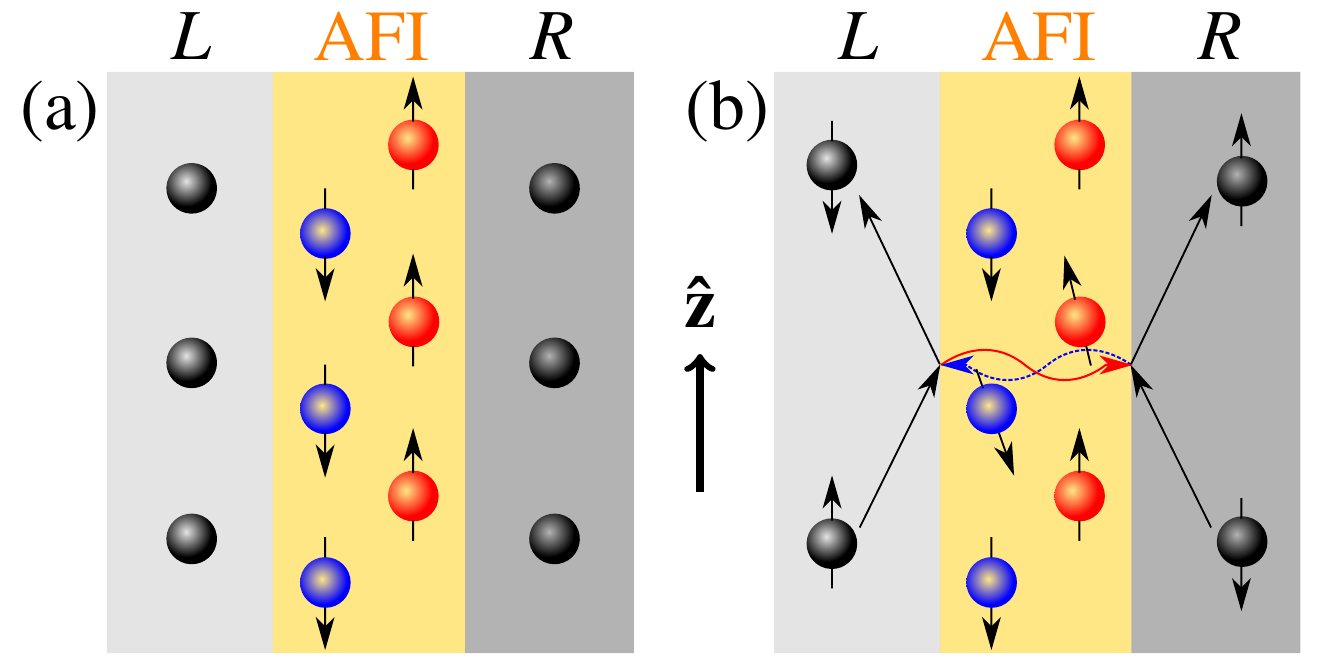}
\caption{(a) An antiferromagnetic insulator (AFI) sandwiched between two separate fermion reservoirs, denoted by $L$ and $R$. We let the spins on sublattice $A$ (illustrated in blue) be down, and the spins on sublattice $B$ (illustrated in red) be up. (b) The fermions in the two reservoirs can interact through emission and absorption of magnons. For the process in the figure we have that either a spin-up fermion in $L$ emits a $S_z=+\hbar$ magnon (red arrow) which is absorbed by a spin-down fermion in $R$, or a spin-down fermion in $R$ emits a $S_z=-\hbar$ magnon (blue dashed arrow) absorbed by a spin-up fermion in $L$.}
\label{fig:System}
\end{figure}

In this Letter, we theoretically demonstrate the application of antiferromagnetic insulators to condensation of indirect excitons.
An exciton is a bound state consisting of an electron and a hole. 
The excitons interact attractively through the Coulomb interaction due to their opposite charges \cite{Wannier1937}. 
Initially predicted many decades ago \cite{Blatt1962,Casella1963}, the exciton condensate has been surprisingly elusive. 
A challenge is that the exciton lifetime is too short to form a condensate due to exciton-exciton annihilation processes such as Auger recombination \cite{OHara:prb:1999,Klimov:science:2000,Wang:prb:2004,Wang:prb:2006}. 
The problem of short exciton lifetimes can be solved by having a spatial separation between the electrons and holes in a trilayer system, where the electrons and holes are separated by an insulating barrier \cite{LOZOVIK1975,LOZOVIK1976,LOZOVIK1977} to drastically lower the recombination rate.
Excitons in such systems are often referred to as (spatially) indirect excitons, and these are ideal to observe the exciton condensate.
Herein, we consider a system where the insulating barrier is an antiferromagnetic insulator, as shown in Fig. \ref{fig:System}.
The insulating barrier can then serve a dual purpose: in addition to increasing the exciton lifetime, the spin fluctuations in the antiferromagnet mediate an additional attractive interaction between the electrons and the holes.
This magnon-mediated interaction cooperates with the Coulomb interaction thereby enabling an increase of the temperature range for observing exciton condensation in experiments.
The exciton lifetimes achieved via antiferromagnetic insulators will be comparable to their nonmagnetic counterparts ($\sim$ \SI{10}{\nano\second} \cite{Calman2018}), leaving the spin-independent physics unaltered.

The indirect exciton condensate has two main experimental signatures. 
The first is a dissipationless counterflow of electric currents in the two layers \cite{Tutuc:prl:2004,Kellogg:prl:2004,Nandi2012}. 
When the exciton condensate moves in one direction, the resulting charge currents in the individual layers are antiparallel due to the oppositely charged carriers in the two layers.
The second signature is a large enhancement of the zero-bias tunneling conductance between the layers \cite{Spielman:prl:2000,Spielman:prl:2001}, reminiscent of the Josephson effect in superconductors.
Comparing the critical condensation temperatures in trilayers with magnetic and nonmagnetic insulating barriers, that otherwise have similar properties and dimensions, should allow to isolate the role of magnons in mediating the condensation.

The exciton condensate is expected to exist when the number of electrons in one layer equals the number of holes in the other. 
Thus far, to the best of our knowledge, experiments with an unequivocal detection of the exciton condensate have utilized quantum Hall systems with a half filling of the lowest Landau level to satisfy this criterion \cite{Eisenstein2004,Wiersma:prl:2004,EisensteinReview2014,Li2017,Liu2017}.
Such systems rely on high external magnetic fields. A recent experiment studying double-bilayer graphene systems has, however, been able to detect the enhanced zero-bias tunneling conductance signature of indirect exciton condensation without any magnetic field, by controlling the electron and hole populations through gate voltages \cite{Tutuc:prl:2018}.
This is an indication of the possible existence of an exciton condensate, and shows promise for finding a magnetic-field free exciton condensate.

In this Letter, we show that the magnon-mediated interaction between the electrons and holes can be attractive or repulsive depending on whether the two magnetic interfaces are with the same or opposite magnetic sublattices. In turn, this enables an unprecedented control over the interaction nature via the variation of the antiferromagnetic insulator thickness by a single atomic layer. Consequently, when the magnon-mediated interaction is paired with the Coulomb interaction, this can be used to control the favored spin structure of the excitons.
In our model, we find that the critical temperature for condensation is enhanced by the exchange interaction in the antiferromagnet, and that a finite magnetic anisotropy is needed to have an attractive interaction around the Fermi level.
Our results suggest that if one lets the insulating barrier in indirect exciton condensation experiments be an antiferromagnetic insulator, the magnon-mediated interactions can significantly strengthen the correlations between the electrons and holes.

\prlsection{Model}
We consider a trilayer system where an antiferromagnetic insulator is sandwiched between two fermion reservoirs, as illustrated in Fig. \ref{fig:System} (a). 
We will then later consider the case where one of these reservoirs is populated by electrons, and the other by holes.
This system can be described by the Hamiltonian $\mathcal{H}=\mathcal{H}_\text{el}+\mathcal{H}_\text{mag}+\mathcal{H}_\text{int}$,
where $\mathcal{H}_\text{el}$ describes the electronic part of the system in the fermion reservoirs, $\mathcal{H}_\text{mag}$ describes the spins in the antiferromagnetic insulator, and $\mathcal{H}_\text{int}$ describes the interfacial interaction between the fermions and magnons.
We assume all three layers to be atomically thin, and thus two-dimensional, for simplicity.

We consider a uniaxial easy-axis antiferromagnetic insulator described by the Hamiltonian
\begin{align}
\mathcal{H}_\text{mag}= J\sum_{\langle i,j\rangle} \ve{S}_i\cdot\ve{S}_j - \frac{K}{2}\sum_{i}S_{iz}^2   \, .
\end{align}
Here $J>0$ is the strength of the nearest-neighbor exchange interaction between the spins which have a magnitude ${\abs{\ve{S}_i}=\hbar S}$ for all $i$, and $K>0$ is the easy-axis anisotropy constant.
Next, we perform a Holstein--Primakoff transformation (HPT) \cite{HolsteinPrimakoff} of the spin operators on each sublattice, denoted by sublattices $A$ and $B$, as defined in Fig. \ref{fig:System}.
From the HPT, we have that the operator $a_i^{\left(\dagger\right)}$ annihilates (creates) a magnon at $\ve{r}_i$ when $\ve{r}_i\in A$, and equivalently $b_i^{\left(\dagger\right)}$ annihilates (creates) a magnon at $\ve{r}_i$ when $\ve{r}_i\in B$.
The magnetic Hamiltonian can be diagonalized through Fourier and Bogoliubov transformations to the form $\mathcal{H}_\text{mag} = \sum_{\ve{k}}\varepsilon_{\ve{k}}\left(\mu_{\ve{k}}^\dagger\mu_{\ve{k}}+\nu_{\ve{k}}^\dagger\nu_{\ve{k}}\right)$.
The magnon energy is given by $\varepsilon_{\ve{k}} = \hbar\sqrt{(1-\gamma_{\ve{k}}^2)\omega_E^2+ \omega_\parallel(2\omega_E+\omega_\parallel)}$, where $\ve{k}$ is the magnon momentum, $\gamma_{\pm\ve{k}}=z^{-1}\sum_{\ve{\delta}}\exp\left(i\ve{k}\cdot\ve{\delta}\right)$, $\ve{\delta}$ a set of vectors to each nearest neighbor, $z$ the number of nearest neighbors, $\omega_E=\hbar J S z$, and $\omega_\parallel = \hbar K S$.
The eigenmagnon operators $\mu_{\ve{k}}^{\left(\dagger\right)}$ and $\nu_{\ve{k}}^{\left(\dagger\right)}$ are related to the HPT magnon operators through the Bogoliubov transformation ${\mu_{\ve{k}}=u_{\ve{k}} a_{\ve{k}}+v_{\ve{k}} b_{-\ve{k}}^\dagger}$, ${\nu_{\ve{k}}=u_{\ve{k}} b_{\ve{k}}+v_{\ve{k}} a_{-\ve{k}}^\dagger}$. The Bogoliubov coefficients $u_{\ve{k}}$ and $v_{\ve{k}}$ are given by $u_{\ve{k}}=\sqrt{(\Gamma_{\ve{k}}+1)/2}$ and $v_{\ve{k}}=\sqrt{(\Gamma_{\ve{k}}-1)/2}$, with ${\Gamma_{\ve{k}}=\{1-[\omega_E\gamma_{\ve{k}}/(\omega_E+\omega_\parallel)]^2\}^{-1/2}}$.

The interfacial exchange interaction between the fermions and magnons at the two magnetic interfaces is modeled by the $s$-$d$ interaction \cite{Zener1951,Kasuya1956}
\begin{align}
\mathcal{H}_\text{int}=-\sum_{j=L,R}\sum_{k=A,B}\sum_{i\in\mathcal{A}_k^j} J_k^j(\ve{r}_i)\hat{\ve{\rho}}_j(\ve{r}_i)\cdot\ve{S}(\ve{r}_i) \, ,
\label{eq:Hint}
\end{align}
which has been successfully applied to describe interactions at magnetic interfaces in similar systems \cite{Takahashi_2010,Kajiwara2010,Zhang:prb:2012,Bender:prb:2015,Kamra:prl:2017}. Here $\mathcal{A}_k^{L(R)}$ is the interface section between the left (right) fermion reservoir and the $k$-th ($k=A,B$) sublattice of the antiferromagnetic insulator. 
The interfacial exchange coupling constants $J_k^j(\ve{r}_i)$ are defined so that they take on the value $J_k^j(\ve{r}_i)=J_k^j$ if $\ve{r}_i\in\mathcal{A}_k^j$, and zero otherwise.
We have also defined the electronic spin density
\begin{align}
\hat{\ve{\rho}}_j(\ve{r}_i) = \frac{1}{2}\sum_{\sigma,\sigma'} \psi_{\sigma,j}^\dagger(\ve{r}_i) \ve{\sigma}_{\sigma\sigma'} \psi_{\sigma',j}(\ve{r}_i)
\end{align}
with $\psi_{\sigma,j}^{\left(\dagger\right)}$ annihilating (creating) a fermion with spin $\sigma$ in the $j$-th ($j=L,R$) fermion reservoir, and $\ve{\sigma}=(\sigma_x,\sigma_y,\sigma_z)$ being a vector of Pauli matrices.

\prlsection{Effective magnon potential}
We will now use a path integral approach where we treat the magnon-fermion interaction as a perturbation, and integrate out the magnonic fields that give rise to processes as illustrated in Fig. \ref{fig:System} (b) to express the interaction as an effective potential between the fermion reservoirs.
We consider the coherent-state path integral $\mathcal{Z}=\int\mathcal{D}^2\psi_L\mathcal{D}^2\psi_R\mathcal{D}^2\mu\mathcal{D}^2\nu \exp\left(-\mathcal{S}/\hbar\right)$
in imaginary time, where $\mathcal{D}^2\mu \equiv \mathcal{D}\mu\mathcal{D}\mu^*$ \textit{etc}. 
The action $\mathcal{S}$ is given by
\begin{align}
    \nonumber \mathcal{S} =& \int_0^{\hbar\beta}\diff\tau \Bigg\{\hbar\sum_i\Bigg[\sum_{\sigma=\uparrow,\downarrow}\sum_{j=L,R} \psi_{\sigma,j}^*(\ve{r}_i,\tau)\partial_\tau\psi_{\sigma,j}(\ve{r}_i,\tau) \\
    &+\sum_{\eta=\mu,\nu} \eta^*(\ve{r}_i,\tau)\partial_\tau\eta(\ve{r}_i,\tau)\Bigg]+\mathcal{H}(\tau)\Bigg\} \, ,
    \label{eq:action}
\end{align}
where $\tau=it$ is imaginary time, and $\beta=1/(k_B T)$ with $k_B$ being the Boltzmann constant and $T$ the temperature.
Note that in the coherent-state path integral we can replace fermion operators by Grassman numbers (\eg $\psi^\dagger\rightarrow\psi^*$) and boson operators by complex numbers (\eg $\eta^\dagger\rightarrow\eta^*$).

We now treat $\mathcal{H}_\text{int}$ as a perturbation, and keep terms up to second order.
We discard any terms that only contribute to intralayer interactions, as we are interested in the interlayer potential between the fermion reservoirs.
By discarding the intralayer terms, we effectively assume that the interlayer interactions will dominate over the intralayer interactions, which is the case for a system designed for indirect exciton condensation.
Next, we integrate out the magnon fields $\mu^{(*)}$ and $\nu^{(*)}$, and write the path integral over the fermion reservoirs as $\mathcal{Z}\approx\int\mathcal{D}^2\psi_L\mathcal{D}^2\psi_R \exp\left(-\mathcal{S}_\text{eff}/\hbar\right)$.
In the momentum and Matsubara-frequency bases, the effective action $\mathcal{S}_\text{eff}$ is given by~\cite{Supp}
\begin{align}
    \nonumber &\mathcal{S}_\text{eff} = \mathcal{S}_\text{el}+\hbar\beta \sum_{\sigma=\uparrow,\downarrow} \sum_{lmn}\sum_{\ve{k}\ve{k'}\ve{q}}U_\sigma(\ve{q},i\omega_n)\psi_{\sigma,L}^*(\ve{k'}+\ve{q},i\nu_l+i\omega_n)\\ 
    &\times \psi_{-\sigma,L}(\ve{k'},i\nu_l)
    \psi_{-\sigma,R}^*(\ve{k}-\ve{q},i\nu_m-i\omega_n)\psi_{\sigma,R}(\ve{k},i\nu_m)\, ,
    \label{eq:S_eff}
\end{align}
where we have here introduced the fermionic and bosonic Matsubara frequencies, $\nu_n=(2n+1)\pi/(\hbar\beta)$ and $\omega_n=2\pi n/(\hbar\beta)$ respectively. 
The action $\mathcal{S}_\text{el}$ describes the contribution of the fermionic fields to the action in Eq. \eqref{eq:action}, except for the contributions from $\mathcal{H}_\text{int}$.
The latter term, $\mathcal{H}_\text{int}$, is instead described by the contribution of the magnon-mediated interlayer-fermion potential
\begin{align}
U_{\sigma}(\ve{q},i\omega_n)\equiv -\frac{\hbar^2 S}{N}\left[\frac{ J_\mu^L(\ve{q})J_\mu^R(\ve{q})}{-\sigma i\hbar\omega_n+\varepsilon_{\ve{q}}} +\frac{ J_\nu^L(\ve{q})J_\nu^R(\ve{q})}{\sigma i\hbar\omega_n+\varepsilon_{\ve{q}}}\right]
\label{eq:general_magnon_potential}
\end{align}
to the effective action, where $N$ is the total number of spin sites in the antiferromagnet.
We assume the two magnetic interfaces are uncompensated, \ie each interface is only with one of the antiferromagnetic sublattices \cite{He2010,Hoogeboom:apl:2017,Kamra:prl:2018} as shown in Fig. \ref{fig:System}.
We compute that the coupling constants $J_{\mu,\nu}^{L,R}(\ve{q})$ describing the effective exchange coupling strength between the spin of the fermions in reservoirs $L$, $R$ to the spin of the eigenmagnons $\mu_{\ve{q}}$, $\nu_{\ve{q}}$ are $J_\mu^{L/R}(\ve{q}) = {v_{\ve{q}}J_B^{L/R}(\ve{r}_{L/R})-u_{\ve{q}} J_A^{L/R}(\ve{r}_{L/R})}$ and $J_\nu^{L/R}(\ve{q}) = {v_{\ve{q}} J_A^{L/R}(\ve{r}_{L/R})-u_{\ve{q}}J_B^{L/R}(\ve{r}_{L/R})}$. 
Since each interface is with only one sublattice, $J_{\mu}^{L}(\ve{q})=-u_{\ve{q}} J_A^{L}$ if the left interface is with sublattice $A$, and  $J_{\mu}^{L}(\ve{q})=v_{\ve{q}} J_B^{L}$ if the left interface is with sublattice $B$. 
We get analogous results for the right interface.
We see that the effective coupling constants $J_{\mu,\nu}^{L,R}(\ve{q})$ can have the same or opposite sign as the coupling constants $J_{A,B}^{L,R}$ depending on which sublattice is at the interface.
This has to do with the spin projection of the eigenmagnon relative to the equilibrium spin direction of the sublattice at the interface.
The effective coupling constants $J_{\mu,\nu}^{L,R}(\ve{q})$ are also enhanced by a Bogoliubov coefficient $u_{\ve{q}}$ or $v_{\ve{q}}$ with respect to the coupling constants $J_{A,B}^{L,R}$.
These are typically large numbers. 
For $\ve{q}=\ve{0}$ we have $u_{\ve{0}}\approx v_{\ve{0}} \approx 2^{-3/4}\times(\omega_E/\omega_\parallel)^{1/4}$ to lowest order in the small ratio $\omega_\parallel/\omega_E$.
The enhancement is due to large spin fluctuations at each sublattice of the antiferromagnet per eigenmagnon in the system, since the eigenmagnons are squeezed states \cite{KamraSqueezing2019,Erlandsen2019}.

By studying Eq. \eqref{eq:general_magnon_potential}, we note that we have $\real[U_{\sigma}(\ve{q},i\omega_n)]<0$ for identical uncompensated interfaces, whereas for a system where one of the interfaces is with sublattice $A$ and the other with sublattice $B$, we have $\real[U_{\sigma}(\ve{q},i\omega_n)]>0$.
Consequentially, this allows us to control whether the magnon-mediated interlayer-fermion potential $U_{\sigma}(\ve{q},i\omega_n)$ is attractive or repulsive by designing the interfaces.
Whether this potential is attractive or repulsive can depend on a single atomic layer.
This allows for an unprecedented high degree of control and tunability of the interlayer-fermion interactions.
The sign difference of the potential can be explained by how the two fermions coupled by the magnon interact with the eigenmagnon spin.
For $\real[U_{\sigma}(\ve{q},i\omega_n)]<0$ we have processes where the fermions couple symmetrically to the magnon spin, \ie both fermions couple either ferromagnetically or antiferromagnetically to its spin. On the other hand, for $\real[U_{\sigma}(\ve{q},i\omega_n)]>0$ we have an asymmetric coupling, where one fermion couples ferromagnetically to the eigenmagnon spin and the other fermion couples antiferromagnetically.

\prlsection{Indirect exciton condensation}
We will now study spontaneous condensation of spatially-indirect excitons where the attraction is mediated by the antiferromagnetic magnons.
We consider the left (right) reservoir to be an n-doped (p-doped) semiconductor.
We describe the semiconductors by the Hamiltonian
\begin{align}
    \mathcal{H}_\text{el} (\tau) = \sum_{j=L,R}\sum_{\ve{k}}\sum_{\sigma=\uparrow,\downarrow}\epsilon_j(\ve{k})\psi_{\sigma,j}^\dagger(\ve{k},\tau)\psi_{\sigma,j}(\ve{k},\tau) \, ,
\end{align}
with $\epsilon_L(\ve{k})=-\epsilon_R(\ve{k})=\hbar^2k^2/(2m)-\epsilon_F\equiv\epsilon(\ve{k})$. 
Here $m$ is the effective electron and hole mass, which we assume to be equal, and $\epsilon_F$ is the Fermi level.
While the operator $\psi_{\sigma,L/R}^\dagger$ creates an electron with spin $\sigma$ in the left/right layer, we note that due to the negative dispersion in the right layer the excitations in this layer are effectively described by electron holes.
We also note that we have not included a Coulomb interaction between the electron and the holes in our model.
The effect of the Coulomb potential on indirect exciton condensation has been widely studied in previous literature \cite{Shevchenko2018}.
We will later argue why the magnon-mediated potential is expected to cooperate with the Coulomb potential in the case of indirect exciton condensation.

The interaction in Eq. \eqref{eq:S_eff} is too complicated for us to solve for the exciton condensation.
We then do an approximation similar to the Bardeen--Cooper--Schrieffer (BCS) theory of superconductivity \cite{BCS,kopnin2001}, and assume that the dominant contribution to the interaction arises when the excitons have zero net momentum ($\ve{k}+\ve{k'}=\ve{q}$), and similarly for the Matsubara frequencies ($i\nu_l+i\nu_m=i\omega_n$).
Next, we introduce the order parameter
\begin{align}
\nonumber\Delta_\sigma(\ve{k},i\nu_m) \equiv& -\sum_n\sum_{\ve{k'}}U_{\sigma}(\ve{k}-\ve{k'},i\nu_m-i\nu_n)\\
&\times\psi^*_{\sigma,R}(\ve{k'},i\nu_n)\psi_{\sigma,L}(\ve{k'},i\nu_n) 
\label{eq:gap}
\end{align}
and its Hermitian conjugate, and perform a Hubbard--Stratonovich transformation of the effective action. 
By doing a saddle-point approximation and integrating over the fermionic fields \cite{Supp}, we then obtain the gap equation 
\begin{align}
\nonumber\Delta_{-\sigma}(\ve{k'},i\nu_n)=&\sum_{m}\sum_{\ve{k}}\beta^{-1}U_{\sigma}(\ve{k}-\ve{k'},i\nu_m-i\nu_n) \\
&\times\frac{\Delta_\sigma(\ve{k},i\nu_m)}{\abs{\Delta_\sigma(\ve{k},i\nu_m)}^2+\epsilon(\ve{k})^2+(\hbar\nu_m)^2} \, .
\label{eq:gap_eqn}
\end{align}
We note that the magnon-mediated potential is attractive when $U_{\sigma}\left(\ve{q},i\omega_n\right)>0$ in the case of indirect exciton condensation, which can be seen from rearranging the fermionic fields in Eq. \eqref{eq:S_eff}.

We now use Eq. \eqref{eq:gap_eqn} to find an analytical expression for the critical temperature $T_c$ below which the excitons spontaneously form a condensate. To obtain an analytical solution, we focus on the case when the gap functions and the magnon-mediated potential are independent of momentum and frequency.
This corresponds to an instantaneous contact interaction, and we therefore assume that the gap functions have an $s$-wave pairing.
Moreover, we see that the gap equation in Eq. \eqref{eq:gap_eqn} only has a solution when $\Delta_\sigma$ and $\Delta_{-\sigma}$ have the same sign.
In the case where spin-degeneracy is unbroken, it is fair to assume that $\Delta_\sigma=\Delta_{-\sigma}$, indicating triplet-like pairing.
In superconductivity, $s$-wave and triplet pairing are mutually exclusive for even frequency order parameters, but in the case of indirect excitons the same symmetry restrictions on the order parameter do not apply, as the composite boson does not consist of identical particles. In other words, for indirect excitons the symmetries in momentum space and spin space are decoupled from one another.
As both the magnon-mediated potential and the Coulomb potential are in the $s$-wave channel and the Coulomb potential is independent of spin, the magnon-mediated potential works together with the Coulomb potential enhancing the attractive exciton pairing interaction.
The fact that we can design whether the magnon-mediated potential is attractive or repulsive allows us to control which spin channel is the most favorable for the excitons to condensate.

\begin{figure}[b]
    \centering
    \includegraphics[width=0.90\linewidth]{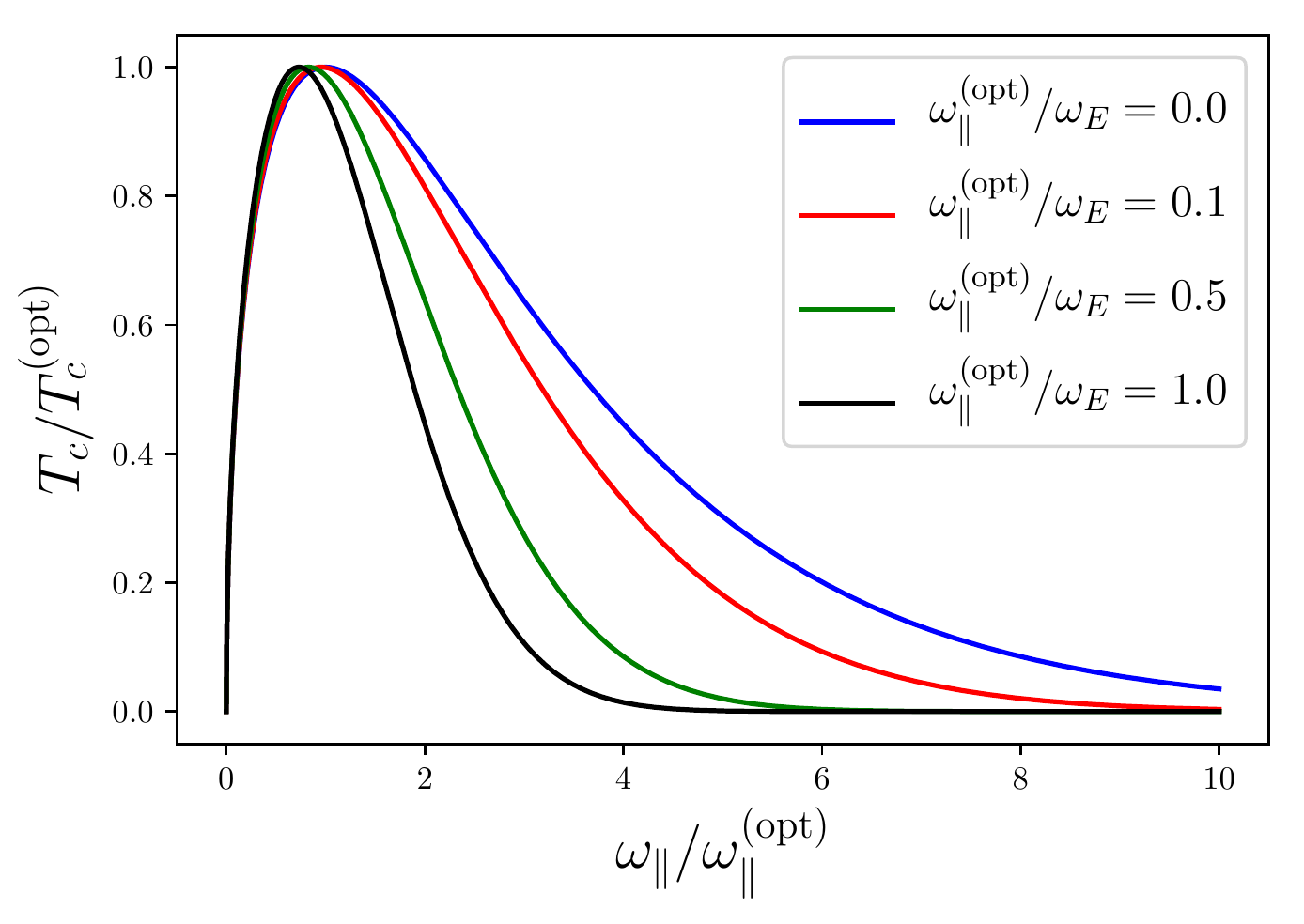}
    \caption{Dependence of the normalized critical temperature on the strength of the normalized magnetic anisotropy.}
    \label{fig:Tc}
\end{figure}

To determine $T_c$ we perform a BCS-like calculation \cite{Supp,kopnin2001} and restrict the sum over Matsubara frequencies to a thin shell around the Fermi level ($\abs{\hbar\nu_m}<\varepsilon_{\ve{0}}$), where the magnon-mediated potential is attractive. 
The analytical expression for $T_c$ is found to be
\begin{align}
 T_c =  \frac{2e^{\gamma_\text{EM}}\varepsilon_{\ve{0}}}{\pi k_B}\exp\left(-\frac{2\pi\varepsilon_{\ve{0}}}{S u_{\ve{0}} v_{\ve{0}} m a^2 J_A^L J_B^R}\right)\, ,
\end{align}
where $\gamma_\text{EM}\approx 0.577$ is the Euler--Mascheroni constant and $a$ the lattice constant of the semiconductors.
Here we have assumed that the left and right magnetic interfaces consist of opposite sublattices. This leads to an attractive exciton interaction.
If we assume the exchange energy among the spins in the bulk is much larger than the interface coupling ($\hbar\omega_E\gg S m a^2 J_A^L J_B^R$), the value of the anisotropy that maximizes $T_c$ is
\begin{align}
\hbar\omega_\parallel^\text{(opt)} \equiv \frac{S m a^2 J_A^L J_B^R}{16\pi} \, .
\end{align}
The full dependence of $T_c$ on the magnetic anisotropy is shown in Fig. \ref{fig:Tc}.
The critical temperature for indirect exciton condensation is largest for a nonzero and finite magnetic anisotropy. This is because in the limit $\omega_\parallel\rightarrow0$ the magnon gap in the antiferromagnetic insulator vanishes, and consequentially so does the thin shell around the Fermi level where the magnon-mediated potential is attractive.
In the case of a large anisotropy, $\omega_\parallel\gg\omega_\parallel^\text{(opt)}$, the enhancement of the magnon-mediated potential due to magnon squeezing is lost \cite{KamraSqueezing2019}.
When the anisotropy takes on its optimal value, the critical temperature becomes
\begin{align}
T_c^{\text{(opt)}}\equiv\frac{\sqrt{\hbar\omega_E S m a^2 J_A^L J_B^R}}{\sqrt{2}\pi^{3/2}k_B}e^{\gamma_\text{EM}-1/2}\, .
\end{align}
Notably, we see that the critical temperature increases monotonously with increasing strength of the exchange interaction $\hbar\omega_E$. The optimal choice of an antiferromagnetic insulator would then be a material with a magnetic anisotropy ($\hbar\omega_\parallel)$ on an energy scale proportional to the exchange coupling at the interface ($\hbar J^{L,R}_{A,B}$), and a very strong exchange interaction in the bulk of the antiferromagnetic insulator ($\hbar\omega_E$). As discussed in the supplemental material~\cite{Supp}, inclusion of retardation and quasiparticle renormalization effects~\cite{McMillan1968,Combescot1990,Marsiglio2018} via Eliashberg method is expected to reduce the $T_c$ estimated here by a factor between $\sqrt{e}$ and $e^{3/2}$. At the same time, accounting for the proper magnon dispersion leads to a similar increase~\cite{Combescot1990} in $T_c$ thereby leaving our estimate essentially unchanged after including these complications.

To show how high the $T_c$ of indirect exciton condensation in our model can be using only the magnon-mediated interaction, we give a numerical estimate for realistic material parameters. 
Using the parameters $S=1$, $m$ equal the electron mass, $a=\SI{5}{\angstrom}$, $\hbar J_A^L=\hbar J_B^R=\SI{10}{\milli\electronvolt}$ \cite{Rohling:prb:2018}, $\omega_E=\SI{8.6d13}{\second^{-1}}$ \cite{Satoh:prl:2010}, and assuming the magnetic anisotropy takes on its optimal value $\omega_\parallel^{\text{(opt)}}=\SI{9.9d9}{\second^{-1}}$, we obtain a $T_c^{\rm (opt)}$ of approximately \SI{7}{\kelvin}.
Antiferromagnetic insulators that can be suitable for the proposed experiment are Cr$_2$O$_3$ \cite{He2010}, $\alpha$-Fe$_2$O$_3$ \cite{Lebrun2018}, and NiO \cite{Satoh:prl:2010}.
A possible emergence of a strong electric field across the barrier could in principle alter the magnetic properties in e.g. Cr$_2$O$_3$ \cite{He2010,Wang2016} and NiO \cite{Lefkidis2007}. We estimate the upper limit of a potential electric field to be around \SI{47}{\volt/\milli\meter}, based on a ``stress-test'' scenario where 1\% of the charge carriers have leaked through the insulating barrier, assuming a charge carrier density of \SI{2.6d10}{\centi\meter^{-2}} \cite{Spielman:prl:2001}. This estimate is considerably weaker than the requirements for influencing typical magnetic insulators \cite{He2010,Wang2016,Lefkidis2007}.
In comparison to the critical temperature above, a recent experiment studying double bilayer graphene in the quantum Hall regime found the Coulomb-mediated exciton condensation to have an activation energy of $\sim\SI{8}{\kelvin}$ \cite{Liu2017}, which was ten times higher than what was found in an experiment using GaAs \cite{Kellogg:prl:2002}.
This demonstrates that the potential mediated by the antiferromagnetic magnons is capable of creating strong correlations between the electrons and holes that could significantly increase the critical temperature for condensation compared to when the excitons are just bound through the Coulomb interaction.

\begin{acknowledgments}
This work was supported by the Research Council of Norway through its Centres of Excellence funding scheme, Project No. 262633 “QuSpin” and Grant No. 239926 “Super Insulator Spintronics,” the European Research Council via Advanced Grant No. 669442 “Insulatronics”, as well as the Stichting voor Fundamenteel Onderzoek der Materie (FOM).
\end{acknowledgments}

\end{document}


\title{Supplemental Material to \\ ``Magnon-Mediated Indirect Exciton Condensation through Antiferromagnetic Insulators''}
\author{Øyvind Johansen}
\email{oyvinjoh@ntnu.no}
\affiliation{Center for Quantum Spintronics, Department of Physics, Norwegian University of Science and Technology, NO-7491 Trondheim, Norway}
\author{Akashdeep Kamra}
\affiliation{Center for Quantum Spintronics, Department of Physics, Norwegian University of Science and Technology, NO-7491 Trondheim, Norway}
\author{Camilo Ulloa}
\affiliation{Institute for Theoretical Physics, Utrecht University, Princetonplein 5,
3584 CC Utrecht, the Netherlands}
\author{Arne Brataas}
\affiliation{Center for Quantum Spintronics, Department of Physics, Norwegian University of Science and Technology, NO-7491 Trondheim, Norway}
\author{Rembert A. Duine}
\affiliation{Center for Quantum Spintronics, Department of Physics, Norwegian University of Science and Technology, NO-7491 Trondheim, Norway}
\affiliation{Institute for Theoretical Physics, Utrecht University, Princetonplein 5,
3584 CC Utrecht, the Netherlands}
\affiliation{Department of Applied Physics, Eindhoven University of Technology, P.O. Box 513, 5600 MB Eindhoven, The Netherlands}
\date{\today}

\maketitle

\section{Interacting system}
In these notes we will consider a trilayer system consisting of an antiferromagnetic insulator  (AFI) sandwiched between two fermion reservoirs (FR|AFI|FR). The magnons in the AFI effectively couple the fermions in the different reservoirs. Here we aim to calculate the magnon-mediated effective interaction between the reservoirs mediated by the magnons by using the path integral formalism. 
\subsection*{One interface}
Let us first analyze a single fermion reservoir placed at the interface of an AFI. We will later generalize our calculation to the trilayer system.
The Hamiltonian consists of three contributions,
\begin{align}
\mathcal{H}=\mathcal{H}_\text{el}+\mathcal{H}_\text{mag}+\mathcal{H}_\text{int} \, ,
\end{align}
where the first two terms describe the fermions in the reservoir and the magnons in the AFI, respectively.
The interfacial coupling between the fermions and magnons is expressed by the $s$-$d$ exchange interaction
\begin{align}
\mathcal{H}_\text{int}=-\sum_{j=A,B}\sum_{i\in\mathcal{A}_j}J_j(\ve{r}_i)\ \hat{\ve{\rho}}(\ve{r}_i)\cdot\ve{S}(\ve{r}_i) \, .
\label{eq:Hint}
\end{align}
Here $\mathcal{A}_{A\, (B)}$ is the cross section of the interface with sublattice $A$ ($B$). The interfacial exchange coupling constant $J_{A/B}(\ve{r}_i)$ is defined as 
\begin{align}
J_j(\ve{r}_i) = 
    \begin{cases}
    J_j,& \text{if } \ve{r}_i\in \mathcal{A}_j\\
    0,              & \text{otherwise} 
\end{cases} \, ,
\label{eq:J_one_interface}
\end{align}
and
\begin{align}
\hat{\ve{\rho}}(\ve{r}_i) = \frac{1}{2}\sum_{\sigma,\sigma'} \psi_\sigma^\dagger(\ve{r}_i) \ve{\sigma}_{\sigma\sigma'} \psi_{\sigma'}(\ve{r}_i)
\label{eq:spin_density}
\end{align}
denotes the spin density. Here $\psi_\sigma^\dagger$ ($\psi_\sigma$) creates (annihilates) a fermion with spin $\sigma$, and $\ve{\sigma}=(\sigma_x,\sigma_y,\sigma_z)$ is a vector of Pauli matrices given by
\begin{align}
\sigma_x=
\begin{pmatrix}
0 & 1\\
1 & 0
\end{pmatrix} \, , \quad
\sigma_y=
\begin{pmatrix}
0 & -i\\
i & 0
\end{pmatrix} \, , \quad
\sigma_z=
\begin{pmatrix}
1 & 0\\
0 & -1
\end{pmatrix} \, .
\end{align}
The direction of the spin $\ve{S}$ appearing in Eq.~\eqref{eq:Hint} depends on the sublattice of the antiferromagnet. We perform a Holstein--Primakoff transformation (HPT) of the spin operators in each sublattice, which for a small number of magnons ($\langle a_i^\dagger a_i\rangle\ , \, \langle b_i^\dagger b_i\rangle\ll 2S$) yields
\begin{subequations}
\begin{align}
S_{i,A}^x&=\frac{\hbar\sqrt{2S}}{2}\left( a_{i}^\dagger+ a_{i}\right) \, ,\quad
S_{i,A}^y=\frac{\hbar\sqrt{2S}}{2i}\left( a_{i}^\dagger- a_{i}\right) \, , \quad
S_{i,A}^z=\hbar\left( a_{i}^\dagger a_{i}-S\right) \, , \label{eq:HP_Transform_A} \\
 S_{i,B}^x&=\frac{\hbar\sqrt{2S}}{2}\left( b_{i}^\dagger+ b_{i}\right) \, ,\quad
S_{i,B}^y=\frac{\hbar\sqrt{2S}}{2i}\left( b_{i}- b_{i}^\dagger\right) \, ,\quad
S_{i,B}^z=\hbar\left(S- b_{i}^\dagger b_{i}\right) \, .
\label{eq:HP_Transform_B}
\end{align}
\label{eq:HP_Transform}
\end{subequations}
We use the HPT in Eq. \eqref{eq:HP_Transform_A} if $\ve{r}_i$ lies in sublattice $A$, and the HPT in Eq. \eqref{eq:HP_Transform_B} if $\ve{r}_i$ lies in sublattice $B$.
The interaction Hamiltonian at the AFI|FR interface can then be written as $\mathcal{H}_\text{int}=\mathcal{H}_\text{int}^A+\mathcal{H}_\text{int}^B$, where
\begin{subequations}
\begin{align}
\mathcal{H}_\text{int}^A &= -J_A\sum_{i\in\mathcal{A}_A} \left[\frac{\hbar\sqrt{2S}}{2}\left(\psi_\uparrow^\dagger\psi_\downarrow a+\psi_\downarrow^\dagger\psi_\uparrow a^\dagger\right)+\frac{\hbar}{2}\left(\psi_\uparrow^\dagger\psi_\uparrow-\psi_\downarrow^\dagger\psi_\downarrow\right)\left( a^\dagger a-S\right)\right] \, , \\
\mathcal{H}_\text{int}^B &= -J_B\sum_{i\in\mathcal{A}_B}\left[\frac{\hbar\sqrt{2S}}{2}\left(\psi_\uparrow^\dagger\psi_\downarrow b^\dagger+\psi_\downarrow^\dagger\psi_\uparrow b\right)-\frac{\hbar}{2}\left(\psi_\uparrow^\dagger\psi_\uparrow-\psi_\downarrow^\dagger\psi_\downarrow\right)\left( b^\dagger b-S\right)\right] \, ,
\end{align}
\end{subequations}
and we are summing over the part of the sublattice cross section $\mathcal{A}_{A/B}$ of sublattice $A/B$ that is in contact with the reservoir of fermions described by $\psi$.
Note that the operators $\psi$, $ a$ and $ b$ have an implicit site index $i$.
The interaction Hamiltonian $\mathcal{H}_\text{int}$ depends on the interface structure of the antiferromagnetic insulator. 
In general, we can express the interaction Hamiltonian as
\begin{align}
\nonumber \mathcal{H}_\text{int} =& -\sum_{i\in(\mathcal{A}_A+\mathcal{A}_B)}\bigg\{\frac{\hbar\sqrt{2S}}{2}\left[\psi_\uparrow^\dagger\psi_\downarrow\left(J_A(\ve{r}_i) a+J_B(\ve{r}_i) b^\dagger\right)\psi_\downarrow^\dagger\psi_\uparrow\left(J_A(\ve{r}_i) a^\dagger+J_B(\ve{r}_i) b\right)\right] \\
&+\frac{\hbar}{2}\left(\psi_\uparrow^\dagger\psi_\uparrow-\psi_\downarrow^\dagger\psi_\downarrow\right)
\times\left[J_A(\ve{r}_i) a^\dagger a-J_B(\ve{r}_i) b^\dagger b+\left(J_B(\ve{r}_i)-J_A(\ve{r}_i)\right)S\right] \bigg\} \, ,
\label{eq:Hint_compensated}
\end{align}
where the sum now runs over the entire reservoir interface.
The interface structure is now encoded in the spatial dependence of the interfacial coupling constants $J_{A,B}(\ve{r}_i)$.
From here on we omit the explicit notation of the $\ve{r}$ dependence of the coupling constants $J_{A,B}(\ve{r}_i)$. In our notation, $J_{A,B}$ depends on $\ve{r}_i$ if it is inside the sum, and it is constant if it is outside the sum.

\section{Path integral}
\label{sec:path_integral}
\subsection{One interface}
Now we calculate the coherent state integral in imaginary time
\begin{align}
\mathcal{Z} = \int \mathcal{D}\psi\mathcal{D}\psi^*\mathcal{D}\mu\mathcal{D}\mu^*\mathcal{D}\nu\mathcal{D}\nu^* \exp\left(-\mathcal{S}/\hbar\right) \, .
\label{eq:pathintegral_one_interface}
\end{align}
where we have introduced the eigenstates $\mu$ and $\nu$ which diagonalize the two-sublattice magnetic Hamiltonian.
The dependence of these eigenstates on the Holstein--Primakoff magnons $a$ and $ b$ will be discussed in the following section.
The action of the bilayer system is $\mathcal{S}=\mathcal{S}_\text{el}+\mathcal{S}_\text{mag}+\mathcal{S}_\text{int}$, with
\begin{subequations}
\begin{align}
\mathcal{S}_\text{el}&=\int_0^{\hbar\beta}\diff \tau\left[\hbar\sum_i\sum_\sigma\psi_\sigma^*(\ve{r}_i,\tau)\dot{\psi}_\sigma(\ve{r}_i,\tau)+\mathcal{H}_\text{el}\right] \, , \\
\label{eq:S_mag} \mathcal{S}_\text{mag}&=\int_0^{\hbar\beta}\diff \tau\left[\hbar\sum_i\sum_{\eta=\mu,\nu}\eta^*(\ve{r}_i,\tau)\dot{\eta}(\ve{r}_i,\tau)+\mathcal{H}_\text{mag}\right] \, , \\
\mathcal{S}_\text{int}&=\int_0^{\hbar\beta}\diff \tau \quad \mathcal{H}_\text{int} \, ,
\end{align}
\end{subequations}
where $\dot{\eta} = \partial\eta/\partial\tau$, $\tau=it$ ($t$ being real time), and $\beta=1/(k_B T)$ with $k_B$ being the Boltzmann constant and $T$ the temperature.
Note that from now on $\psi^{\left(*\right)}$ are Grassman variables such that $\psi^{\left(\dagger\right)}=\psi^{\left(*\right)}$, and $\lbrace \mu^{\left(*\right)},\ \nu^{\left(*\right)},\  a^{\left(*\right)},\ b^{\left(*\right)}\rbrace$ are complex numbers.
We treat the interaction term as a perturbation and perform an expansion up to second order in the interaction term of the action
\begin{align}
 \mathcal{Z}&=\int \mathcal{D}^2\psi\mathcal{D}^2\mu\mathcal{D}^2\nu \exp\left( -\frac{\mathcal{S}_\text{el}+\mathcal{S}_\text{mag}}{\hbar}\right) \exp\left(-\frac{\mathcal{S}_\text{int}}{\hbar}\right) \approx \int \mathcal{D}^2\psi\mathcal{D}^2\mu\mathcal{D}^2\nu \exp\left( -\frac{\mathcal{S}_\text{el}+\mathcal{S}_\text{mag}}{\hbar}\right)\gamma_{\rm int} \label{eq:path_integral_gammaint}\,
\end{align}
where $\gamma_{\rm int} = 1- \mathcal{S}_{\rm int}/\hbar + (\mathcal{S}_{\rm int}/\hbar)^2/2$.
We from now on also use the shorthand notation $\mathcal{D}^2\psi=\mathcal{D}\psi\mathcal{D}\psi^*$ \textit{etc}.
The first order term in $\gamma_\text{int}$ that is linear in $\mathcal{S}_\text{int}$ is neglected, as all the processes in Eq. \eqref{eq:Hint_compensated} are only between a single fermion and a magnon. Such processes to first order can therefore not mediate an interaction between two separate fermions, which are the processes we are interested in later on.

\begin{figure}[tpb!]
\centering
\begin{tikzpicture}
\node[above right] (img) at (0,0) {\includegraphics[width=0.9\linewidth]{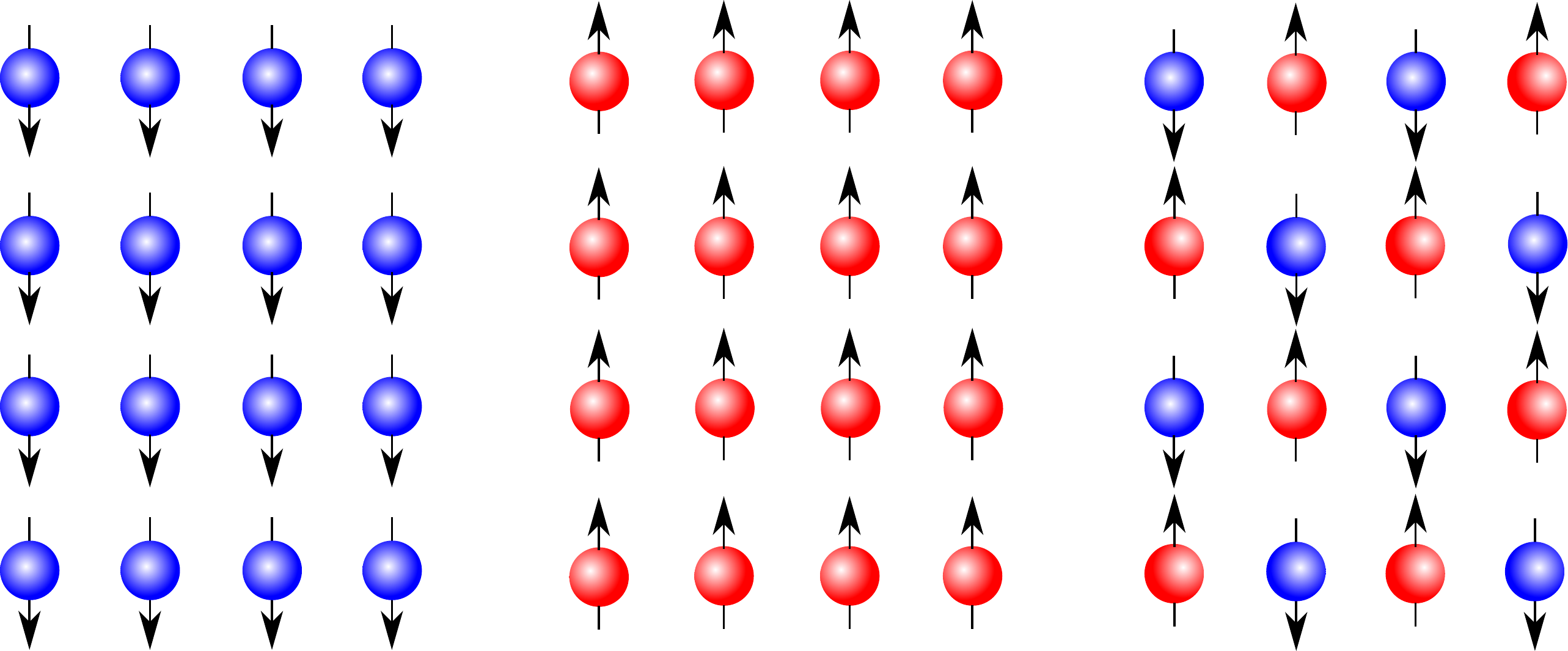}};
\node at (-10pt,170pt) {\Large{(a)}};
\node at (153pt,170pt) {\Large{(b)}};
\node at (323pt,170pt) {\Large{(c)}};
\end{tikzpicture}
\caption{The spin structures of the atoms in the AFI in the plane closest to the fermion reservoir as seen from the fermion reservoir. (a) and (b) are fully uncompensated interfaces with sublattices $A$ and $B$ respectively, where only one of the sublattices is present at the interface. (c) is a compensated interface where both sublattices are present at the interface, in a chessboard pattern.}
\label{fig:Interfaces}
\end{figure}

Let us now analyze the second order term, and evaluate
\begin{align}
\mathcal{S}_\text{int}^2 = \int \diff \tau \diff \tau' \mathcal{H}_\text{int}\left(\tau\right)\mathcal{H}_\text{int}\left(\tau'\right) \, .
\end{align}
We write $\psi_\sigma\equiv \psi_\sigma(\tau,\ve{r}_i)$ and $\underline{\psi}_\sigma\equiv \psi_\sigma(\tau',\ve{r}_j)$, and the same convention for $ a$, $b$, and $J_{A,B}(\ve{r}_i)$. Let us first consider a general type of interface, and use $\mathcal{H}_\text{int}$ in \eqref{eq:Hint_compensated}. The interface structure is then encoded in the spatial dependencies of $J_{A,B}(\ve{r}_i)$. Note that the two limits of a fully uncompensated interface (either with sublattice $A$ or sublattice $B$, as defined in Fig. \ref{fig:Interfaces}) can easily be recovered from the general expression by setting either $J_A$ or $J_B$ to zero.
We then find
\begin{align}
\nonumber\mathcal{H}_\text{int}(\tau)\mathcal{H}_\text{int}(\tau') = \sum_{i,j\in(\mathcal{A}_A+\mathcal{A}_B)} \bigg[\frac{\hbar^2 S}{2}&\left(A_{AA} J_A\underline{J}_A+A_{AB} J_A \underline{J}_B + A_{BA} J_B \underline{J}_A + A_{BB} J_B\underline{J}_B\right) \\
+\frac{\hbar^2}{4}&\left(B_{AA} J_A\underline{J}_A+B_{AB} J_A \underline{J}_B  
+ B_{BA} J_B \underline{J}_A + B_{BB} J_B\underline{J}_B\right)\bigg] \, ,
\end{align}
where we have defined
\begin{widetext}
\begin{subequations}
\begin{align}
A_{AA} &= \psi_\uparrow^\dagger\psi_\downarrow\underline{\psi}_\uparrow^\dagger\underline{\psi}_\downarrow a\underline{ a} + \psi_\uparrow^\dagger\psi_\downarrow\underline{\psi}_\downarrow^\dagger\underline{\psi}_\uparrow a\underline{ a}^\dagger + \psi_\downarrow^\dagger\psi_\uparrow\underline{\psi}_\uparrow^\dagger\underline{\psi}_\downarrow a^\dagger\underline{ a}+ \psi_\downarrow^\dagger\psi_\uparrow\underline{\psi}_\downarrow^\dagger\underline{\psi}_\uparrow a^\dagger\underline{ a}^\dagger \, , \\
A_{AB} &= \psi_\uparrow^\dagger\psi_\downarrow\underline{\psi}_\uparrow^\dagger\underline{\psi}_\downarrow  a\underline{ b}^\dagger + \psi_\uparrow^\dagger\psi_\downarrow\underline{\psi}_\downarrow^\dagger\underline{\psi}_\uparrow  a\underline{ b} + \psi_\downarrow^\dagger\psi_\uparrow\underline{\psi}_\uparrow^\dagger\underline{\psi}_\downarrow  a^\dagger\underline{ b}^\dagger + \psi_\downarrow^\dagger\psi_\uparrow\underline{\psi}_\downarrow^\dagger\underline{\psi}_\uparrow  a^\dagger\underline{ b} \, , \\
A_{BA} &= \psi_\uparrow^\dagger\psi_\downarrow\underline{\psi}_\uparrow^\dagger\underline{\psi}_\downarrow  b^\dagger\underline{ a} + \psi_\uparrow^\dagger\psi_\downarrow\underline{\psi}_\downarrow^\dagger\underline{\psi}_\uparrow  b^\dagger\underline{ a}^\dagger
+ \psi_\downarrow^\dagger\psi_\uparrow\underline{\psi}_\uparrow^\dagger\underline{\psi}_\downarrow  b\underline{ a} + \psi_\downarrow^\dagger\psi_\uparrow\underline{\psi}_\downarrow^\dagger\underline{\psi}_\uparrow  b\underline{ a}^\dagger \, , \\
A_{BB} &= \psi_\uparrow^\dagger\psi_\downarrow\underline{\psi}_\uparrow^\dagger\underline{\psi}_\downarrow b^\dagger\underline{ b}^\dagger + \psi_\uparrow^\dagger\psi_\downarrow\underline{\psi}_\downarrow^\dagger\underline{\psi}_\uparrow b^\dagger\underline{ b} + \psi_\downarrow^\dagger\psi_\uparrow\underline{\psi}_\uparrow^\dagger\underline{\psi}_\downarrow b\underline{ b}^\dagger+ \psi_\downarrow^\dagger\psi_\uparrow\underline{\psi}_\downarrow^\dagger\underline{\psi}_\uparrow b\underline{ b} \, , \\
B_{AA} &= \left(\psi_\uparrow^\dagger\psi_\uparrow\underline{\psi}_\uparrow^\dagger\underline{\psi}_\uparrow-\psi_\uparrow^\dagger\psi_\uparrow\underline{\psi}_\downarrow^\dagger\underline{\psi}_\downarrow-\psi_\downarrow^\dagger\psi_\downarrow\underline{\psi}_\uparrow^\dagger\underline{\psi}_\uparrow+\psi_\downarrow^\dagger\psi_\downarrow\underline{\psi}_\downarrow^\dagger\underline{\psi}_\downarrow\right) \times\left( a^\dagger a\underline{ a}^\dagger\underline{ a}-S  a^\dagger a -S \underline{ a}^\dagger\underline{ a} +S^2\right) \\
B_{AB} &= \left(\psi_\uparrow^\dagger\psi_\uparrow\underline{\psi}_\uparrow^\dagger\underline{\psi}_\uparrow-\psi_\uparrow^\dagger\psi_\uparrow\underline{\psi}_\downarrow^\dagger\underline{\psi}_\downarrow-\psi_\downarrow^\dagger\psi_\downarrow\underline{\psi}_\uparrow^\dagger\underline{\psi}_\uparrow+\psi_\downarrow^\dagger\psi_\downarrow\underline{\psi}_\downarrow^\dagger\underline{\psi}_\downarrow\right)\times \left(- a^\dagger a\underline{ b}^\dagger\underline{ b}+S a^\dagger a + S\underline{ b}^\dagger\underline{ b}-S^2\right) \\
B_{BA} &= \left(\psi_\uparrow^\dagger\psi_\uparrow\underline{\psi}_\uparrow^\dagger\underline{\psi}_\uparrow-\psi_\uparrow^\dagger\psi_\uparrow\underline{\psi}_\downarrow^\dagger\underline{\psi}_\downarrow-\psi_\downarrow^\dagger\psi_\downarrow\underline{\psi}_\uparrow^\dagger\underline{\psi}_\uparrow+\psi_\downarrow^\dagger\psi_\downarrow\underline{\psi}_\downarrow^\dagger\underline{\psi}_\downarrow\right)\times \left(- b^\dagger b\underline{ a}^\dagger\underline{ a} + S\underline{ a}^\dagger\underline{ a} +S b^\dagger b -S^2\right) \\
B_{BB} &= \left(\psi_\uparrow^\dagger\psi_\uparrow\underline{\psi}_\uparrow^\dagger\underline{\psi}_\uparrow-\psi_\uparrow^\dagger\psi_\uparrow\underline{\psi}_\downarrow^\dagger\underline{\psi}_\downarrow-\psi_\downarrow^\dagger\psi_\downarrow\underline{\psi}_\uparrow^\dagger\underline{\psi}_\uparrow+\psi_\downarrow^\dagger\psi_\downarrow\underline{\psi}_\downarrow^\dagger\underline{\psi}_\downarrow\right) \times\left( b^\dagger b\underline{ b}^\dagger\underline{ b}-S  b^\dagger b -S \underline{ b}^\dagger\underline{ b} +S^2\right) \, .
\end{align}
\label{eq:one_interface_coeffs}
\end{subequations}
\end{widetext}

The contributions from terms that are odd in the number of magnon operators are neglected because they correspond to disconnected diagrams whose expectation values will vanish. 
The terms $A_{AB}$, $A_{BA}$ are not discarded as $a$ and $ b$ are not eigenexcitations, and the product of the two might still contain contributing terms that are even in the eigenmagnon operators.
From this point on we also disregard terms of order $a^4$ and $b^4$ and higher to be consistent with the linear expansion of the HPT.
Moreover, the interactions proportional to $ a a$, $ a b^\dagger$, $ b b$, and their Hermitian conjugates vanish once the magnon operators are diagonalized. This is because these processes do not conserve spin.

\subsection{Two interfaces}
We now want to generalize to the case where we have two fermion reservoirs $L$ and $R$ connected through the magnetic insulator.
Similar to the case with only one fermion reservoir, there is no contribution of interest from the first order perturbation in $\mathcal{S}_\text{int}$ as these processes can not mediate an interaction between the two fermion reservoirs.
Moreover, the processes represented by $B_{ij}$ ($i,j=A,B$) for the one-interface case in Eq. \eqref{eq:one_interface_coeffs} also cannot couple the fermion reservoirs.
This is because there are no processes in $B_{ij}$ where a magnon is emitted at one point and absorbed at another, only processes where a magnon is instantaneously emitted and absorbed at the same point.
These interactions (first order in $\mathcal{S}_\text{int}$ and processes proportional to $B_{ij}$) will only contribute to a renormalization of the energy at each interface.
As we are only interested in contributions that can mediate a coupling between the two reservoirs, we henceforth drop all contributions that are first order in $\mathcal{S}_\text{int}$ or included in $B_{ij}$.

The second order contribution can be expressed as
\begin{align}
\left(\frac{\mathcal{S}_\text{int}}{\hbar}\right)^2 = \int \diff \tau \diff \tau' \left\{\sum_{\substack{i,j=L,R\\i\neq j}}\sum_{k,l=A,B}\sum_{\substack{m\in\mathcal{A}_i\\n\in\mathcal{A}_j}} J_k^i(\ve{r}_m)\underline{J}_l^j(\ve{r}_n)\left[\hat{\ve{\rho}}_i(\ve{r}_m,\tau)\cdot\ve{S}(\ve{r}_m,\tau)\right] \left[\hat{\underline{\ve{\rho}}}_j(\ve{r}_n,\tau')\cdot\underline{\ve{S}}(\ve{r}_n,\tau')\right] \right\} \, ,
\end{align}
with $\mathcal{A}_{L(R)}=\mathcal{A}_A^{L(R)}+\mathcal{A}_B^{L(R)}$, and $\hat{\ve{\rho}}_i$ is the spin density defined in Eq. \eqref{eq:spin_density} in the $i$-th ($i=L,R$) fermion reservoir.
For consistency with the contributions we have previously neglected, we only sum over the contributions where the fermions are located in different layers ($i\neq j$).
We also allow for the interfaces to be different, and have generalized the definition of the interfacial exchange couplings in Eq. \eqref{eq:J_one_interface} to
\begin{align}
J_k^i(\ve{r}_m) = 
    \begin{cases}
    J_k^i,& \text{if } \ve{r}_m\in \mathcal{A}_k^i\\
    0,& \text{otherwise} 
\end{cases} \, .
\label{eq:J_two_interfaces}
\end{align}
From this point on, we only consider uncompensated magnetic interfaces, corresponding to either Fig. \ref{fig:Interfaces} (a) or (b).
In other words, if we have the left interface being an uncompensated interface with sublattice $A$, we have $J_A ^L(\ve{r}) = J_A^L$ for all $\ve{r}\in\mathcal{A}_L$ as well as $J_B ^L(\ve{r}) = 0$ for all $\ve{r}\in\mathcal{A}_L$.
However, we make no assumptions about which sublattice is at the interface, and whether the two interfaces are with identical or opposite sublattices.

We see that the result for one interface can easily be generalized to two interfaces with different fermion reservoirs.
We can define
\begin{align}
\gamma_\text{int}^{(2)}=1+\frac{S}{4}\int\diff \tau\diff \tau' \sum_{\substack{i,j=L,R\\i\neq j}}\sum_{k,l=A,B}\sum_{\substack{m\in\mathcal{A}_i\\n\in\mathcal{A}_j}}  A_{kl}^{ij}J_k^i(\ve{r}_m)\underline{J}_l^j(\ve{r}_n) \, ,
\label{eq:gamma2int}
\end{align}
where the coefficients $A_{kl}^{ij}$ are as in the one-reservoir case, but now with two different reservoir labels. As an example, we have
\begin{align}
 A_{AA}^{LR} &= \psi_{\uparrow,L}^\dagger\psi_{\downarrow,L}\underline{\psi}_{\uparrow,R}^\dagger\underline{\psi}_{\downarrow,R} a\underline{ a} + \psi_{\uparrow,L}^\dagger\psi_{\downarrow,L}\underline{\psi}_{\downarrow,R}^\dagger\underline{\psi}_{\uparrow,R} a\underline{ a}^\dagger + \psi_{\downarrow,L}^\dagger\psi_{\uparrow,L}\underline{\psi}_{\uparrow,R}^\dagger\underline{\psi}_{\downarrow,R} a^\dagger\underline{ a}+ \psi_{\downarrow,L}^\dagger\psi_{\uparrow,L}\underline{\psi}_{\downarrow,R}^\dagger\underline{\psi}_{\uparrow,R} a^\dagger\underline{ a}^\dagger \, ,
\end{align}
where $\psi_{\uparrow,L}^\dagger$ creates a fermion with spin up in the left fermion reservoir, and so on.
$\gamma_\text{int}^{(2)}$ takes the role of $\gamma_\text{int}$ in the path integral in Eq. \eqref{eq:path_integral_gammaint} for the two-reservoir case.
This path integral is also extended by an integration over the fermionic fields in the second fermion reservoir, \ie $\mathcal{D}^2\psi\rightarrow\mathcal{D}^2\psi_L\mathcal{D}^2\psi_R$.

\section{Magnetic Hamiltonian}
Now that we have considered how the fermions interact with the magnons in the HPT-magnon basis, we wish to find how these $a$ and $b$ magnons relate to the eigenexcitations of the system (the $\mu$ and $\nu$ magnons, which we integrate over in the path integral in Eq. \eqref{eq:pathintegral_one_interface}).
We consider an easy-axis antiferromagnetic insulator, which is described by the Hamiltonian
\begin{align}
\mathcal{H}_\text{mag}= J\sum_{\langle i,j\rangle} \ve{S}_i\cdot\ve{S}_j - \frac{K}{2}\sum_{i}S_{iz}^2   \, .
\end{align}
where $J$ and $K$ are the strengths of the exchange interaction and magnetic anisotropy, respetively. Performing a HPT of the spin operators as defined in Eq. \eqref{eq:HP_Transform}, disregarding any constant terms in the Hamiltonian and only keeping terms to second order in the magnon-operators, we find
\begin{align}
\mathcal{H}_\text{mag}=&\frac{J\hbar^2 S}{2}\sum_{\ve{\delta}}\left[\sum_{i\in A}\left( a_i^\dagger a_i+ b_{i+\ve{\delta}}^\dagger b_{i+\ve{\delta}}+2 a_i^\dagger b_{i+\ve{\delta}}^\dagger\right) +\sum_{i\in B}\left( b_i^\dagger b_i+ a_{i+\ve{\delta}}^\dagger a_{i+\ve{\delta}}+2 b_i a_{i+\ve{\delta}}\right)\right] +\hbar^2 K S\left(\sum_{i\in A} a_i^\dagger a_i+\sum_{i\in B} b_i^\dagger b_i\right) \, .
\end{align}
Here $\ve{\delta}$ is a set of nearest-neighbor vectors from site $i$.
Next, we perform a Fourier transformation of the magnon operators, given by
\begin{subequations}
\label{eq:magnon_FT}
\begin{align}
 a(\ve{r}_i,\tau) &= \frac{1}{\sqrt{N_A}}\sum_{\ve{k}}  a_{\ve{k}}(\tau) e^{i\ve{k}\cdot\ve{r}_i} \, , \quad
 a^\dagger(\ve{r}_i,\tau) = \frac{1}{\sqrt{N_A}}\sum_{\ve{k}}  a_{\ve{k}}^\dagger(\tau) e^{-i\ve{k}\cdot\ve{r}_i}  \, , \\
 b(\ve{r}_i,\tau) &= \frac{1}{\sqrt{N_B}}\sum_{\ve{k}}  b_{\ve{k}}(\tau) e^{i\ve{k}\cdot\ve{r}_i} \, , \quad
 b^\dagger(\ve{r}_i,\tau) = \frac{1}{\sqrt{N_B}}\sum_{\ve{k}}  b_{\ve{k}}^\dagger(\tau) e^{-i\ve{k}\cdot\ve{r}_i}  \, .
\end{align}
\end{subequations}
where $N_{A/B}$ is the number of spins in sublattice $A/B$. The momentum $\ve{k}$ in each sum runs over the sublattice Brillouin zone. For an antiferromagnet, we have $N_A=N_B$.

If we assume that $N_A$ is macroscopic, the terms in the Hamiltonian transform as
\begin{align}
\sum_{j\in A}  a^\dagger (\ve{r}_j,\tau) a (\ve{r}_j,\tau) = \sum_{\ve{k}} a_{\ve{k}}^\dagger(\tau) a_{\ve{k}}(\tau) \, .
\end{align}
Transforming the remaining terms in the Hamiltonian to momentum space, we find it to be
\begin{align}
\mathcal{H}_\text{mag}=& \sum_{\ve{k}} \Big\{\hbar^2S\left(Jz+K\right)\left[ a_{\ve{k}}^\dagger(\tau) a_{\ve{k}}(\tau)+ b_{\ve{k}}^\dagger(\tau) b_{\ve{k}}(\tau)\right]+\hbar^2SJz\left[\gamma_{\ve{k}} a_{\ve{k}}^\dagger(\tau) b_{-\ve{k}}^\dagger(\tau)+\gamma_{-\ve{k}} a_{\ve{k}}(\tau) b_{-\ve{k}}(\tau)\right]\Big\} \, ,
\label{eq:Hmag_HPT_Fourier}
\end{align}
where $z$ is the number of nearest neighbours, and
\begin{align}
\gamma_{\ve{k}}=z^{-1}\sum_{\ve{\delta}}e^{i\ve{k}\cdot\ve{\delta}} = \gamma_{-\ve{k}} \, .
\end{align}
We now want to diagonalize the Hamiltonian in Eq. \eqref{eq:Hmag_HPT_Fourier}. We use a Bogoliubov transformation given by
\begin{align}
\mu_{\ve{k}}(\tau) = u_{\ve{k}}  a_{\ve{k}}(\tau)+v_{\ve{k}} b_{-\ve{k}}^\dagger(\tau) \, , \quad
\nu_{\ve{k}}(\tau) = u_{\ve{k}}  b_{\ve{k}}(\tau)+v_{\ve{k}} a_{-\ve{k}}^\dagger(\tau) \, ,
\end{align}
where the new bosonic operators $\mu$ and $\nu$ also satisfy bosonic commutation relations.
Introducing the quantities $\omega_E=J\hbar S z$ and $\omega_\parallel= K\hbar S$, we find that
\begin{align}
\label{eq:HAFI_diagonal}
\mathcal{H}_\text{mag}= \sum_{\ve{k}}  \left[\varepsilon_{\ve{k},\mu}\mu_{\ve{k}}^\dagger(\tau)\mu_{\ve{k}}(\tau)+\varepsilon_{\ve{k},\nu}\nu_{\ve{k}}^\dagger(\tau)\nu_{\ve{k}}(\tau)\right]\, .
\end{align}
The energies are
\begin{align}
\varepsilon_{\ve{k},\mu}=\varepsilon_{\ve{k},\nu}\equiv\varepsilon_{\ve{k}}=\hbar\sqrt{\left(1-\gamma_{\ve{k}}^2\right)\omega_E^2+\omega_\parallel\left(2\omega_E+\omega_\parallel\right)} \, .
\end{align}
The Bogoliubov coefficients are
\begin{align}
u_{\ve{k}} = \sqrt{\frac{\Gamma_{\ve{k}}+1}{2}} \quad \, , \quad v_{\ve{k}} = \sqrt{\frac{\Gamma_{\ve{k}}-1}{2}} \, ,
\end{align}
where we have introduced
\begin{align}
\Gamma_{\ve{k}} = \frac{1}{\sqrt{1-\left(\frac{\omega_E \gamma_{\ve{k}}}{\omega_E+\omega_\parallel}\right)^2}} \, .
\end{align}

\section{Magnon Green's functions}
Now that we can express the fermion-magnon interaction in terms of the eigenmagnons $\mu$ and $\nu$, over which we integrate, we can proceed by performing the integrals over the magnon fields to express the interaction as an effective fermion-fermion interaction mediated by the magnons.
Let us introduce the (inverse) Green's function so that the partition function for the magnons becomes
\begin{align}
&\mathcal{Z}_\text{mag} = \int \mathcal{D}^2\mu\mathcal{D}^2\nu \exp\left[\int_0^{\hbar\beta}\diff\tau\diff\tau'\sum_{i,j} \ve{\Phi}^\dagger (\ve{r}_i,\tau) \mathcal{G}_\text{mag}^{-1}(\ve{r}_i,\tau ;\ve{r}_j,\tau')\ve{\Phi} (\ve{r}_j,\tau') \right] \, ,
\end{align}
where we have
\begin{align}
\ve{\Phi}^\dagger (\ve{r},\tau) = \left(\mu^*(\ve{r},\tau),\quad \nu^*(\ve{r},\tau)\right) \, .
\end{align}
To describe the (imaginary) time dependence of the magnon fields we do a Matsubara expansion and go to frequency space,
\begin{align}
\mu(\ve{r},\tau)&=\frac{1}{\sqrt{N_A}}\sum_{n=-\infty}^\infty\sum_{\ve{k}}  \mu_{\ve{k},n}e^{i(\ve{k}\cdot\ve{r}-\omega_n\tau)} \, , 
\end{align}
where $\omega_n=2\pi n/(\hbar\beta)$ is the Matsubara frequency for bosons.

In a momentum and frequency representation, the partition function can alternatively be expressed as
\begin{align}
\mathcal{Z}_\text{mag} &= \int \mathcal{D}^2\mu\mathcal{D}^2\nu \exp\left(-\frac{\mathcal{S}_\text{mag}}{\hbar}\right) = \int \mathcal{D}^2\mu\mathcal{D}^2\nu \exp\left[\sum_{\ve{k},\ve{k'}} \sum_{n,n'} \ve{\Phi}^\dagger_{\ve{k},n} \mathcal{G}_\text{mag}^{-1}(\ve{k},i\omega_n ;\ve{k'},i\omega_{n'})\ve{\Phi}_{\ve{k'},n'} \right] \, ,
\end{align}
with
\begin{align}
\ve{\Phi}^\dagger_{\ve{k},n} = \left(\mu^*_{\ve{k},n},\quad \nu^*_{\ve{k},n}\right) \, .
\end{align}
Performing the Matsubara expansion of our fields, and using our Hamiltonian as shown in Eq. \eqref{eq:HAFI_diagonal} in terms of the Matsubara modes, we note that the action in Eq. \eqref{eq:S_mag} becomes
\begin{align}
\mathcal{S}_\text{mag} = \hbar\beta\sum_{\ve{k}} \sum_{n} \sum_{\eta=\mu,\nu}\left(-i\hbar\omega_n+\varepsilon_{\ve{k},\eta}\right)\eta_{\ve{k},n}^*\eta_{\ve{k},n} \, ,
\end{align}
where we have used the identity
\begin{align}
\int_0^{\hbar\beta}\diff\tau \frac{e^{i(\omega_{n'}-\omega_n)\tau}}{\hbar\beta} = \delta_{n,n'} \, .
\end{align}
Consequentially, we find the Green's function in the Matsubara basis to be
\begin{align}
\mathcal{G}_\text{mag}(\ve{k},i\omega_n ;\ve{k'},i\omega_{n'}) &= -\frac{\hbar}{\hbar\beta}\delta_{\ve{k}\ve{k'}}\delta_{n,n'}\begin{pmatrix}
\left(-i\hbar\omega_n+\varepsilon_{\ve{k},\mu}\right)^{-1} & 0 \\
0 & \left(-i\hbar\omega_n+\varepsilon_{\ve{k},\nu}\right)^{-1}
\end{pmatrix} =
-\begin{pmatrix}
\langle\mu_{\ve{k},n}^*\mu_{\ve{k'},n'}\rangle & 0 \\
0 & \langle\nu_{\ve{k},n}^*\nu_{\ve{k'},n'}\rangle
\end{pmatrix} \, .
\end{align}

Now we wish to calculate expectation values such as
\begin{align}
\langle\mu^* (\ve{r},\tau)\mu (\ve{r'},\tau')\rangle = \mathcal{Z}_\text{mag}^{-1}\int \mathcal{D}^2\mu\mathcal{D}^2\nu \exp\left(-\frac{\mathcal{S}_\text{mag}}{\hbar}\right)\mu^* (\ve{r},\tau)\mu (\ve{r'},\tau') \, .
\end{align}
We once again do a Fourier transform and a Matsubara expansion, and using the results above we find that
\begin{align}
\nonumber \langle\mu^* (\ve{r},\tau)\mu (\ve{r'},\tau')\rangle &= \frac{1}{N_A}\sum_{\ve{k},\ve{k'}} \sum_{n,n'} e^{i(\ve{k'}\cdot\ve{r'}-\ve{k}\cdot\ve{r})}e^{i(\omega_n\tau-\omega_{n'}\tau')} \mathcal{Z}_\text{mag}^{-1}\int \mathcal{D}^2\mu\mathcal{D}^2\nu \exp\left(-\frac{\mathcal{S}_\text{mag}}{\hbar}\right)\mu^*_{\ve{k},n}\mu_{\ve{k'},n'} \\
\nonumber &= \frac{1}{N_A}\sum_{\ve{k},\ve{k'}}\sum_{n,n'} e^{i(\ve{k'}\cdot\ve{r'}-\ve{k}\cdot\ve{r})}e^{i(\omega_n\tau-\omega_{n'}\tau')} \langle\mu_{\ve{k},n}^*\mu_{\ve{k'},n'}\rangle \\
\nonumber &= \frac{1}{\hbar\beta N_A}\sum_{\ve{k}}\sum_{n}\frac{\hbar}{-i\hbar\omega_n+\varepsilon_{\ve{k},\mu}}e^{i\ve{k}\cdot(\ve{r'}-\ve{r})}e^{-i\omega_n(\tau'-\tau)} \\
&\equiv \sum_{\ve{k}} \langle\mu^*_{\ve{k}} (\ve{r},\tau)\mu_{\ve{k}} (\ve{r'},\tau')\rangle = -\mathcal{G}_\text{mag}(\ve{r},\tau ;\ve{r'},\tau')\, .
\end{align}
A similar result is obtained for $\langle\nu^* (\ve{r},\tau)\nu (\ve{r'},\tau')\rangle$, and all other expectation values vanish.

Previously we expressed the interaction in terms of the sublattice magnon operators $ a$ and $ b$.
The expectation values of these magnons are related to the expectation values of the diagonal magnons by the following:
\begin{subequations}
\begin{align}
\langle a^*(\ve{r},\tau) a(\ve{r'},\tau')\rangle 
= &\sum_{\ve{k}} \Big[u_{\ve{k}}^2\langle\mu_{\ve{k}}^*(\ve{r},\tau)\mu_{\ve{k}}(\ve{r'},\tau')\rangle +v_{\ve{k}}^2\langle\nu_{\ve{k}}^*(\ve{r'},\tau')\nu_{\ve{k}}(\ve{r},\tau)\rangle\Big] \, , \\
\langle b^*(\ve{r},\tau) b(\ve{r'},\tau')\rangle 
= &\sum_{\ve{k}} \Big[ v_{\ve{k}}^2\langle\mu_{\ve{k}}^*(\ve{r'},\tau')\mu_{\ve{k}}(\ve{r},\tau)\rangle +u_{\ve{k}}^2\langle\nu_{\ve{k}}^*(\ve{r},\tau)\nu_{\ve{k}}(\ve{r'},\tau')\rangle\Big] \, , \\
\langle a(\ve{r},\tau) b(\ve{r'},\tau')\rangle 
= &-\sum_{\ve{k}}  u_{\ve{k}} v_{\ve{k}}\Big[\langle\mu_{\ve{k}}^*(\ve{r'},\tau')\mu_{\ve{k}}(\ve{r},\tau)\rangle +\langle\nu_{\ve{k}}^*(\ve{r},\tau)\nu_{\ve{k}}(\ve{r'},\tau')\rangle\Big] \, , \\
\langle a^*(\ve{r},\tau) b^*(\ve{r'},\tau')\rangle 
= &-\sum_{\ve{k}} u_{\ve{k}} v_{\ve{k}}\Big[\langle\mu_{\ve{k}}^*(\ve{r},\tau)\mu_{\ve{k}}(\ve{r'},\tau')\rangle+\langle\nu_{\ve{k}}^*(\ve{r'},\tau')\nu_{\ve{k}}(\ve{r},\tau)\rangle\Big] \, .
\end{align}
\end{subequations}
The other expectation values vanish.

\section{Effective potential}
Using the results found previously, we integrate out the magnons and find that
\begin{align}
\mathcal{Z} &= \mathcal{Z}_\text{mag} \int\mathcal{D}^2\psi_L\mathcal{D}^2\psi_R \exp{\left(-\mathcal{S}_\text{el}^\text{eff}/\hbar\right)} \equiv \int\mathcal{D}^2\psi_L\mathcal{D}^2\psi_R \exp{\left(-\mathcal{S}_\text{el}/\hbar\right)} 
\int\mathcal{D}^2\mu\mathcal{D}^2\nu \exp(-\mathcal{S}_\text{mag}/\hbar)\gamma_\text{int}^{(2)} \, ,
\end{align}
where $\gamma_\text{int}^{(2)}$ is defined in Eq. \eqref{eq:gamma2int}, and we have defined the effective action of the fermionic system:
\begin{align}
\nonumber\mathcal{S}_{\text{el}}^\text{eff} &= \int_0^{\hbar\beta}\diff\tau\left[\hbar\sum_{\substack{i=L,R}}\sum_{\ve{r}_j\in i}\sum_{\sigma=\uparrow,\downarrow} \psi_{\sigma,i}^*(\ve{r}_j,\tau)\dot{\psi}_{\sigma,i}(\ve{r}_j,\tau)+\mathcal{H}_\text{el}\right]+\hbar\left(1-\gamma_\text{int}^{G}\right) \, .
\end{align}
We have here reintroduced the interaction term in the exponent.

The interaction term can be expressed as following:
\begin{align}
\hbar\left( 1-\gamma_\text{int}^{G}\right) =& \sum_{\substack{i,j=L,R\\i\neq j}}\int\diff\tau\diff\tau'\sum_{\substack{k\in\mathcal{A}_i\\l\in\mathcal{A}_j}} \sum_{\sigma=\uparrow,\downarrow}V_{\sigma,-\sigma}^{ij}\left(\ve{r}_k,\ve{r}_l,\tau-\tau'\right)\psi_{\sigma,i}^*(\ve{r}_k,\tau)\psi_{-\sigma,i}(\ve{r}_k,\tau) \psi_{-\sigma,j}^*(\ve{r}_l,\tau')\psi_{\sigma,j}(\ve{r}_l,\tau')  \, .
\end{align}
The effective potential $V_{\sigma,-\sigma}^{ij}(\ve{r},\ve{r'},\tau-\tau')$ describes a spin-flip interaction of two fermions located in reservoirs $i$ and $j$ mediated by a magnon in the magnetic insulator.
One fermion flips its spin and thereby emits a magnon. This magnon is then absorbed by a fermion that is located in the reservoir on the opposing side of the insulating barrier with respect to the fermion that emitted the magnon.
We then write down the effective spin-flip potential as
\begin{subequations}
\begin{align}
&V_{\uparrow,\downarrow}^{ij}\left(\ve{r},\ve{r'},\tau-\tau'\right) = V_{\downarrow,\uparrow}^{ji}\left(\ve{r'},\ve{r},\tau'-\tau\right) \\
\nonumber =& -\frac{\hbar S}{4}\Bigg[\langle a^*(\ve{r'},\tau') a(\ve{r},\tau)\rangle J_A^i(\ve{r})J_A^j(\ve{r'})+\langle a(\ve{r},\tau) b(\ve{r'},\tau')\rangle J_A^i(\ve{r})J_B^j(\ve{r'}) \\
&+ \langle a^*(\ve{r'},\tau') b^*(\ve{r},\tau)\rangle J_A^i(\ve{r'})J_B^j(\ve{r}) + \langle b^*(\ve{r},\tau) b(\ve{r'},\tau')\rangle J_B^i(\ve{r})J_B^j(\ve{r'}) \Bigg] \\
\nonumber =& -\frac{ S}{4 \beta N_A}\sum_{\ve{k}}\sum_{n}\frac{\hbar}{-i\hbar\omega_n+\varepsilon_{\ve{k},\mu}}\left[v_{\ve{k}}J_B^i(\ve{r})-u_{\ve{k}} J_A^i(\ve{r})\right]\left[v_{\ve{k}}J_B^j(\ve{r'})-u_{\ve{k}} J_A^j(\ve{r'})\right]e^{i\ve{k}\cdot(\ve{r}-\ve{r'})}e^{-i\omega_n(\tau-\tau')} \\
&- \frac{S}{4\beta N_A}\sum_{\ve{k}}\sum_{n}\frac{\hbar}{-i\hbar\omega_n+\varepsilon_{\ve{k},\nu}}\left[v_{\ve{k}} J_A^i(\ve{r})-u_{\ve{k}}J_B^i(\ve{r})\right]\left[v_{\ve{k}} J_A^j(\ve{r'})-u_{\ve{k}}J_B^j(\ve{r'})\right]e^{i\ve{k}\cdot(\ve{r'}-\ve{r})}e^{-i\omega_n(\tau'-\tau)}
\end{align}
\end{subequations}

We now do a Fourier transformation and Matsubara expansion of the fermionic fields, so that
\begin{align}
\psi_{\sigma,i}(\ve{r},\tau) = \frac{1}{\sqrt{N_i}}\sum_n\sum_{\ve{k}}\psi_{\sigma,i}(\ve{k},i\nu_n)e^{i\ve{k}\cdot\ve{r}-i\nu_n\tau} \, .
\end{align}
Here $N_i$ is the number of sites in reservoir $i$ ($i=L,R$), and $\ve{k}$ now runs over the Brillouin zone of the fermion reservoirs.
The Matsubara expansion of the fermionic fields is defined in terms of the fermionic Matsubara frequencies $\nu_n=(2n+1)\pi/(\hbar\beta)$.
With these transformations, we rewrite the magnon-mediated interaction as
\begin{align}
 \nonumber&\sum_{\substack{i,j=L,R\\i\neq j}}\sum_{\sigma=\uparrow,\downarrow}\int\diff\tau\diff\tau'\sum_{\substack{k\in\mathcal{A}_i\\l\in\mathcal{A}_j}}V_{\sigma,-\sigma}^{ij}\left(\ve{r}_k,\ve{r}_l,\tau-\tau'\right) \psi_{\sigma,i}^*(\ve{r}_k,\tau)\psi_{-\sigma,i}(\ve{r}_k,\tau) \psi_{-\sigma,j}^*(\ve{r}_l,\tau')\psi_{\sigma,j}(\ve{r}_l,\tau')  \\
 =&\sum_{\substack{i,j=L,R\\i\neq j}}\sum_{\sigma=\uparrow,\downarrow}\left(\hbar\beta\right)^2 \sum_{lmn}\sum_{\ve{k}\ve{k'}\ve{q}}V_{\sigma,\mu\nu}^{ij}\left(\ve{q},i\omega_n\right) \psi_{\sigma,i}^*(\ve{k'}+\ve{q},i\nu_l+i\omega_n)\psi_{-\sigma,i}(\ve{k'},i\nu_l) \psi_{-\sigma,j}^*(\ve{k}-\ve{q},i\nu_m-i\omega_n)\psi_{\sigma,j}(\ve{k},i\nu_m)  \, .
\label{eq:Seff_LR_sigma}
\end{align}
Here we have defined
\begin{align}
V_{\sigma,-\sigma}^{ij}(\ve{r},\ve{r'},\tau-\tau') \equiv \sum_{\ve{q}}\sum_{n} V_{\sigma,\mu\nu}^{ij}(\ve{q},i\omega_n) e^{i\ve{q}\cdot(\ve{r}-\ve{r'})}e^{-i\omega_n(\tau-\tau')} \, ,
\end{align}
with
\begin{align}
\nonumber V_{\sigma,\mu\nu}^{ij}(\ve{q},i\omega_n)\equiv -\frac{\hbar S}{4 \hbar\beta N_A}
\Bigg\{&\frac{\hbar}{-\sigma i\hbar\omega_n+\varepsilon_{\ve{q},\mu}}\left[v_{\ve{q}}J_B^i(\ve{r}_i)-u_{\ve{q}} J_A^i(\ve{r}_i)\right]\left[v_{\ve{q}}J_B^j(\ve{r}_j)-u_{\ve{q}} J_A^j(\ve{r}_j)\right] \\
+&\frac{\hbar}{\sigma i\hbar\omega_n+\varepsilon_{\ve{q},\nu}}\left[v_{\ve{q}} J_A^i(\ve{r}_i)-u_{\ve{q}}J_B^i(\ve{r}_i)\right]\left[v_{\ve{q}} J_A^j(\ve{r}_j)-u_{\ve{q}}J_B^j(\ve{r}_j)\right]\Bigg\} \, .
\label{eq:general_magnon_potential}
\end{align}
While at first glance the potential in the above equation seemingly also depends on position, we note that since we are only considering uncompensated interfaces (see Fig. \ref{fig:Interfaces} (a) and (b)) the interfacial coupling constants $J_{A,B}^{L,R}(\ve{r})$ can only take on the constant values $J_{A,B}^{L,R}$ or zero for all positions $\ve{r}$ at each interface.
As an example, $J_A^L(\ve{r}_L\in\mathcal{A}_L)=J_A^L$ if the interface with reservoir $L$ is an uncompensated interface with sublattice $A$, but $J_A^L(\ve{r}_L\in\mathcal{A}_L)=0$ if the interface with reservoir $L$ is an uncompensated interface with sublattice $B$.
However, if one is considering a compensated interface such as in Fig. \ref{fig:Interfaces} (c), the interfacial coupling constants have a periodic spatial dependence at that interface, and they would then also need to be Fourier transformed accordingly.
Since we are only considering uncompensated interfaces, we do not have to take this into consideration.

We have previously assumed an inversion symmetry in the magnetic insulator, so that $\varepsilon_{\ve{k},\mu/\nu} = \varepsilon_{-\ve{k},\mu/\nu}$. With this in mind, the Fourier transform of the effective magnon potential has the following symmetries:
\begin{align}
V_{\sigma,\mu\nu}^{ij}(\ve{k},i\omega_n) = V_{\sigma,\mu\nu}^{ji}(\ve{k},i\omega_n) \, , \quad V_{\sigma,\mu\nu}^{ij}(\ve{k},i\omega_n) = V_{\sigma,\mu\nu}^{ij}(-\ve{k},i\omega_n) \, , \quad V_{-\sigma,\mu\nu}^{ij}(\ve{k},i\omega_n) = V_{\sigma,\mu\nu}^{ij}(\ve{k},-i\omega_n) \, .
\end{align}
Through these symmetries, and some relabelling of the momenta and Matsubara frequencies, one can see that the interactions in \eqref{eq:Seff_LR_sigma} are pairwise identical. The interactions are paired through the substitutions $L \leftrightarrow R$, $\sigma\rightarrow -\sigma$.
Consequently, we can perform the sums over layers (only considering interlayer interactions) in \eqref{eq:Seff_LR_sigma}, and reduce the result to
\begin{align}
&2\left(\hbar\beta\right)^2 \sum_{\sigma=\uparrow,\downarrow}\sum_{lmn}\sum_{\ve{k}\ve{k'}\ve{q}} 
V_{\sigma,\mu\nu}^{LR}\left(\ve{q},i\omega_n\right)\psi_{\sigma,L}^*(\ve{k'}+\ve{q},i\nu_l+i\omega_n) \psi_{-\sigma,L}(\ve{k'},i\nu_l)\psi_{-\sigma,R}^*(\ve{k}-\ve{q},i\nu_m-i\omega_n)\psi_{\sigma,R}(\ve{k},i\nu_m)  \, .
\end{align}
To have a more intuitive form of the magnon-mediated potential, we define the quantity
\begin{align}
    U_{\sigma}\left(\ve{q},i\omega_n\right) \equiv 2\hbar\beta V_{\sigma,\mu\nu}^{LR}\left(\ve{q},i\omega_n\right) \, ,
\end{align}
which has units Joule.

\section{Exciton interaction}
We will now use the effective action to look at interlayer exciton condensation. It is then advantageous to reorder the fermionic fields on the following form:
\begin{align}
 &-\hbar\beta \sum_{\sigma=\uparrow,\downarrow}\sum_{lmn}\sum_{\ve{k}\ve{k'}\ve{q}} U_{\sigma}\left(\ve{q},i\omega_n\right)\psi_{\sigma,L}^*(\ve{k'}+\ve{q},i\nu_l+i\omega_n)\psi_{\sigma,R}(\ve{k},i\nu_m)\psi_{-\sigma,R}^*(\ve{k}-\ve{q},i\nu_m-i\omega_n)\psi_{-\sigma,L}(\ve{k'},i\nu_l)  \, .
\end{align}
We can simplify this further by only considering the case where the exciton has a net zero momentum ($\ve{k'}+\ve{q}-\ve{k}=0$). Moreover, we assume that the electronic fields in an exciton pair has the same Matsubara frequency ($i\nu_l+i\omega_n=i\nu_m$) as well as momentum. The simplified interaction we consider is then
\begin{align}
-\hbar\beta \sum_{mn}\sum_{\ve{k}\ve{k'}}\sum_\sigma U_{\sigma}\left(\ve{k}-\ve{k'},i\nu_m-i\nu_n\right)\psi_{\sigma,L}^*(\ve{k},i\nu_m) \psi_{\sigma,R}(\ve{k},i\nu_m)\psi_{-\sigma,R}^*(\ve{k'},i\nu_n)\psi_{-\sigma,L}(\ve{k'},i\nu_n)  \, .
\end{align}

\section{Hubbard--Stratonovich transformation}
To simplify the Hamiltonian, we can reduce the effective electronic Hamiltonian to a bilinear Hamiltonian by introducing an auxiliary Hubbard--Stratonovich field.
To introduce the new Hubbard--Stratonovich fields, we first multiply our path integral by a unity path integral, given by the path integral over a white-noise field $\alpha$ \cite{coleman_2015}:
\begin{align}
\mathcal{Z}_\alpha =& \int\mathcal{D}^2\alpha \exp\Bigg[-\sum_\sigma\sum_{mn}\sum_{\ve{k}\ve{k'}}\alpha_{\sigma}^\dagger(\ve{k},i\nu_m)\beta U_{\sigma}^{-1}(\ve{k}-\ve{k'},i\nu_m-i\nu_n)\alpha_{-\sigma}(\ve{k'},i\nu_n)\Bigg] \, .
\end{align}
Defining the bilinears $A_\sigma(\ve{k},i\nu_n) =\psi_{\sigma,R}^*(\ve{k},i\nu_n)\psi_{\sigma,L}(\ve{k},i\nu_n)$ and its Hermitian conjugate in the fermionic fields, we can shift the white-noise field variables by introducing the Hubbard--Stratonovich field
\begin{align}
\Delta_\sigma(\ve{k},i\nu_m) = \alpha_\sigma(\ve{k},i\nu_m)-\sum_n\sum_{\ve{k'}}U_{\sigma}(\ve{k}-\ve{k'},i\nu_m-i\nu_n)A_\sigma(\ve{k'},i\nu_n)\, ,
\end{align}
and its Hermitian conjugate.
Noting that the inverse  matrix $U^{-1}_{\sigma}(\ve{k}-\ve{k'},i\nu_m-i\nu_n)$ is defined so that it satisfies the relation
\begin{align}
\sum_l\sum_{\ve{k}}U^{-1}_{\sigma}(\ve{k}-\ve{k'},i\nu_l-i\nu_m)U_{\sigma}(\ve{k}-\ve{k''},i\nu_l-i\nu_n) = \delta_{\ve{k'},\ve{k''}}\delta_{m,n} \, ,
\label{eq:inverse_relation}
\end{align}
we can rewrite the sum of the interaction and white-noise fields in the effective action as
\begin{align}
\nonumber&-\hbar\beta\sum_{mn}\sum_{\ve{k}\ve{k'}} A_{\sigma}^\dagger(\ve{k},i\nu_m)U_{\sigma}(\ve{k}-\ve{k'})A_{-\sigma}(\ve{k'},i\nu_n) \\
\nonumber &+ \hbar\beta\sum_{mn}\sum_{\ve{k}\ve{k'}}\alpha_{\sigma}^\dagger(\ve{k},i\nu_m)U_{\sigma}^{-1}(\ve{k}-\ve{k'},i\nu_m-i\nu_n)\alpha_{-\sigma}(\ve{k'},i\nu_n) \\
\nonumber &= \hbar\beta\sum_n\sum_{\ve{k}}\Delta^\dagger_{\sigma}(\ve{k},i\nu_n)A_{-\sigma}(\ve{k},i\nu_n) + \hbar\beta\sum_n\sum_{\ve{k}}A^\dagger_{\sigma}(\ve{k},i\nu_n)\Delta_{-\sigma}(\ve{k},i\nu_n) \\
&+ \hbar\beta\sum_{mn}\sum_{\ve{k}\ve{k'}}\Delta_{\sigma}^\dagger(\ve{k},i\nu_m)U_{\sigma}^{-1}(\ve{k}-\ve{k'},i\nu_m-i\nu_n)\Delta_{-\sigma}(\ve{k'},i\nu_n) \, .
\end{align}
The path integral in its entirety can then be written as
\begin{align}
\mathcal{Z} =& \int\mathcal{D}^2\Delta\exp\left[-\beta\sum_\sigma\sum_{mn}\sum_{\ve{k}\ve{k'}}\Delta_{\sigma}^\dagger(\ve{k},i\nu_m) U_{\sigma}^{-1}(\ve{k}-\ve{k'},i\nu_m-i\nu_n)\Delta_{-\sigma}(\ve{k'},i\nu_n)\right]\int\mathcal{D}^2\psi_L\int\mathcal{D}^2\psi_R\exp\left(-\frac{\tilde{S}_\text{eff}}{\hbar}\right) \, ,
\end{align}
where we have defined
\begin{align}
\tilde{S}_\text{eff} =& \int_0^{\hbar\beta}\diff\tau\left[\hbar\sum_{i=L,R}\sum_{\sigma=\uparrow,\downarrow}\sum_{j\in V_i} \psi_{\sigma,i}^*(\ve{r}_j,\tau)\dot{\psi}_{\sigma,i}(\ve{r}_j,\tau)+\mathcal{H}_\text{el}\right] + \hbar\beta\sum_\sigma\sum_n\sum_{\ve{k}}\left[\Delta^\dagger_{\sigma}(\ve{k})A_{-\sigma}(\ve{k},i\nu_n) + A^\dagger_{\sigma}(\ve{k},i\nu_n)\Delta_{-\sigma}(\ve{k})\right] \, .
\end{align}
Assuming the electron and hole Hamiltonians to be diagonalized on the form
\begin{align}
\mathcal{H}_\text{el} = \sum_n\sum_{\ve{k}}\epsilon(\ve{k})\sum_\sigma\Big[ \psi_{\sigma,L}^*(\ve{k},i\nu_n)\psi_{\sigma,L}(\ve{k},i\nu_n) 
- \psi_{\sigma,R}^*(\ve{k},i\nu_n)\psi_{\sigma,R}(\ve{k},i\nu_n)\Big] \, ,
\end{align}
we can do Fourier and Matsubara expansions of the kinetic term to write the effective electronic action as
\begin{align}
\nonumber \tilde{S}_\text{eff} =& \hbar\beta\sum_{\sigma,n,{\ve{k}}}\Big\{ \psi_{\sigma,L}^*(\ve{k},i\nu_n)\left[-i\hbar\nu_n + \epsilon(\ve{k})\right]\psi_{\sigma,L}(\ve{k},i\nu_n) +  \psi_{\sigma,R}^*(\ve{k},i\nu_n)\left[-i\hbar\nu_n - \epsilon(\ve{k})\right]\psi_{\sigma,R}(\ve{k},i\nu_n) \\
&+ \Delta^\dagger_{\sigma}(\ve{k},i\nu_n)\psi_{-\sigma,R}^*(\ve{k},i\nu_n)\psi_{-\sigma,L}(\ve{k},i\nu_n) + \psi_{\sigma,L}^*(\ve{k},i\nu_n)\psi_{\sigma,R}(\ve{k},i\nu_n)\Delta_{-\sigma}(\ve{k},i\nu_n)\Big\} \, .
\end{align}
Here $\epsilon(\ve{k})$ is the energy of the fermions in the left reservoir as a function of momentum, and the negative energy of the fermions in the right reservoir.

\section{Saddle-point approximation}
Under the path integral, the Hubbard--Stratonovich transformation is exact. We have now successfully made our action bilinear in the fermionic fields, but this came at the cost of introducing another path integral over a bosonic Hubbard--Stratonovich field. 
To simplify our calculations, we make a mean-field approximation where we assume that the path integral over the Hubbard--Stratonovich field is dominated by the values of $\Delta_\sigma(\ve{k})$ and $\Delta_\sigma^*(\ve{k})$ that minimize the total action
\begin{align}
 S_\Delta =& \hbar\beta\sum_\sigma\sum_{mn}\sum_{\ve{k}\ve{k'}}\Delta_{-\sigma}^\dagger(\ve{k},i\nu_m) U_{\sigma}^{-1}(\ve{k}-\ve{k'},i\nu_m-i\nu_n)\Delta_{\sigma}(\ve{k'},i\nu_n) + \tilde{S}_\text{eff} \, .
\end{align}
The path integral is then approximated by
\begin{align}
 \mathcal{Z}\approx & \exp\left[-\beta\sum_\sigma\sum_{mn}\sum_{\ve{k}\ve{k'}}\Delta_{-\sigma}^*(\ve{k},i\nu_m) U_{\sigma}^{-1}(\ve{k}-\ve{k'},i\nu_m-i\nu_n)\Delta_{\sigma}(\ve{k'},i\nu_n)\right]\int\mathcal{D}^2\psi_L\int\mathcal{D}^2\psi_R\exp\left(-\frac{\tilde{S}_\text{eff}}{\hbar}\right) \, ,
\end{align}
where we the Hubbard--Stratonovich fields are no longer integrated over, but now take on constant values that satisfy $\delta S_\Delta/\delta[\Delta_\sigma^{(*)}(\ve{k},i\nu_m)]=0$. This is known as the saddle-point approximation.

\section{Gap equation}
The effective action $\tilde{S}_\text{eff}$ now only consists of bilinear terms in the fermionic fields, and can therefore be written on the form
\begin{align}
\tilde{S}_\mathrm{eff} &= \hbar\beta\sum_n\sum_{\ve{k}}
\ve{\Psi}^\dagger(\ve{k},i\nu_n)
\left[-i\hbar\nu_n\underline{1} + \underline{h}_\Psi(\ve{k},i\nu_n)\right]
\ve{\Psi}(\ve{k},i\nu_n) \, ,
\end{align}
where $\underline{1}$ is the identity matrix, and we have defined
\begin{align}
\underline{h}_\Psi(\ve{k},i\nu_n) \equiv 
\begin{pmatrix}
\epsilon(\ve{k}) & \Delta_\uparrow(\ve{k},i\nu_n) & 0 & 0 \\
\Delta_\uparrow^*(\ve{k},i\nu_n) & -\epsilon(\ve{k}) & 0 & 0 \\
0 & 0 & -\epsilon(\ve{k}) & \Delta_\downarrow^*(\ve{k},i\nu_n) \\
0 & 0 & \Delta_\downarrow(\ve{k},i\nu_n) & \epsilon(\ve{k})
\end{pmatrix} \, .
\end{align}
$\ve{\Psi}(\ve{k},i\nu_n)$ is the extended Nambu spinor $\ve{\Psi}(\ve{k},i\nu_n) = (\psi_{\downarrow,L}(\ve{k},i\nu_n),\, \psi_{\downarrow,R}(\ve{k},i\nu_n),\, \psi_{\uparrow,R}(\ve{k},i\nu_n),\, \psi_{\uparrow,L}(\ve{k},i\nu_n))^T$.
Performing the Gaussian integral over the fermionic fields, we can express the effective action as
\begin{align}
\nonumber \frac{S_\Delta}{\hbar} =& - \sum_{\ve{k}}\sum_n\mathrm{ln}\left\{\det \left[\beta(-i\hbar\nu_n\underline{1} + \underline{h}_\Psi(\ve{k},i\nu_n))\right]\right\} \\
\nonumber &+ \beta\sum_\sigma\sum_{mn}\sum_{\ve{k}\ve{k'}}\Delta_{\sigma}^*(\ve{k},i\nu_m)U_{\sigma}^{-1}(\ve{k}-\ve{k'},i\nu_m-i\nu_n)\Delta_{-\sigma}(\ve{k'},i\nu_n) \\
\nonumber =& - \sum_{\sigma}\sum_{n}\sum_{\ve{k}}\mathrm{ln}\left\{\beta^2\left[\abs{\Delta_\sigma(\ve{k},i\nu_n)}^2+\epsilon(\ve{k})^2+(\hbar\nu_n)^2\right]\right\} \\
&+ \beta\sum_\sigma\sum_{mn}\sum_{\ve{k}\ve{k'}}\Delta_{\sigma}^*(\ve{k},i\nu_m)U_{\sigma}^{-1}(\ve{k}-\ve{k'},i\nu_m-i\nu_n)\Delta_{-\sigma}(\ve{k'},i\nu_n)  \, .
\end{align}
Imposing the saddle-point approximation condition that $\delta S_\Delta/\delta[\Delta_\sigma^{(*)}(\ve{k},i\nu_m)]=0$, we find the gap equation to become
\begin{align}
\frac{\delta S_\Delta}{\delta \left[\Delta_\sigma^*(\ve{k},i\nu_m)\right]} =& -\frac{\Delta_\sigma(\ve{k},i\nu_m)}{\abs{\Delta_\sigma(\ve{k},i\nu_m)}^2+\epsilon(\ve{k})^2+(\hbar\nu_m)^2} + \beta\sum_n\sum_{\ve{k'}}U_{\sigma}^{-1}(\ve{k}-\ve{k'},i\nu_m-i\nu_n)\Delta_{-\sigma}(\ve{k'},i\nu_n) = 0 \, .
\end{align}
Inverting the above gap equation through the identity in Eq. \eqref{eq:inverse_relation}, we obtain
\begin{align}
\Delta_{-\sigma}(\ve{k'},i\nu_n)=&\sum_{m}\sum_{\ve{k}}\beta^{-1}U_{\sigma}(\ve{k}-\ve{k'},i\nu_m-i\nu_n) \frac{\Delta_\sigma(\ve{k},i\nu_m)}{\abs{\Delta_\sigma(\ve{k},i\nu_m)}^2+\epsilon(\ve{k})^2+(\hbar\nu_m)^2} \, .
\label{eq:gap_eqn_matsubara_momentum}
\end{align}
Solving the general case for when the gap depends on momentum and frequency is extremely challenging, as one then has to solve for an infinite set of coupled self-consistent equations. We can reduce the complexity by only considering frequency independent gaps. This assumption requires that only the frequency independent part of the magnon-mediated potential contributes to the exciton pairing. We can then consider the potential and gaps as constants when performing the Matsubara sum.

Defining $E_\sigma(\ve{k})\equiv \sqrt{\abs{\Delta_{\sigma}(\ve{k})}^2+\left[\epsilon(\ve{k})\right]^2}$, we note that
\begin{align}
\sum_m \frac{1}{E_{\sigma}(\ve{k})^2+\left(\hbar\nu_m\right)^2} = \sum_m\sum_\pm \frac{1}{2E_{\sigma}(\ve{k})} \frac{\pm 1}{i\hbar\nu_m\pm E_{\sigma}(\ve{k})} \, .
\end{align}
Using the Matsubara sum
\begin{align}
\frac{1}{\hbar\beta}\sum_m\frac{1}{i\nu_m-E/\hbar} = \frac{1}{e^{\beta E}+1} \equiv n_F(E) \, ,
\end{align}
we perform the Matsubara sum in the gap equation:
\begin{align}
\sum_m \frac{1}{E_{\sigma}(\ve{k})^2+\left(\hbar\nu_m\right)^2} = \frac{\beta}{2E_{\sigma}(\ve{k})}\left[n_F(-E_{\sigma}(\ve{k}))-n_F(E_{\sigma}(\ve{k}))\right] = \frac{\beta}{2E_{\sigma}(\ve{k})}\tanh\left[\frac{\beta E_{\sigma}(\ve{k})}{2}\right] \, . 
\end{align}
The gap equation is then simplified to
\begin{align}
\Delta_{-\sigma}(\ve{k'})=\sum_{\ve{k}}U_{\sigma}(\ve{k}-\ve{k'},0)\frac{\Delta_{\sigma}(\ve{k})}{2E_{\sigma}(\ve{k})}\tanh\left[\frac{\beta E_{\sigma}(\ve{k})}{2}\right] \, . 
\label{eq:gap_eqn_momentum}
\end{align}
Note that the exciton interaction is attractive when ${U_{\sigma}(\ve{k}-\ve{k'},0)>0}$, which explains the sign difference from the typical form of the normal BCS gap equation.

\section{Critical temperature}
We now wish to determine an analytical estimate of the critical temperature of the exciton condensation. Finding an analytical self-consistent solution of Eq. \eqref{eq:gap_eqn_momentum} is too complicated, so we will focus on the simple limit where both the magnon-mediated potential and the gap functions are independent of both momentum and frequency.
Starting from Eq. \eqref{eq:gap_eqn_matsubara_momentum}, we then approximate the gap equation as
\begin{align}
\Delta_{-\sigma}=&\beta^{-1}U_{\sigma}(\ve{0},0)\Delta_{\sigma}\frac{N_A}{(2\pi/a)^2} \sum_{n}\int\diff^2k\frac{1}{\abs{\Delta_{\sigma}}^2+\left[\epsilon(\ve{k})\right]^2+\left(\hbar\nu_n\right)^2} \, .
\end{align}
Here we have gone to the continuum limit:
\begin{align}
\frac{1}{N_{L/R}}\sum_{\ve{k}} \rightarrow \frac{1}{A_k}\int\diff^2k \, ,
\end{align}
where we assume that $N_{L/R}=N_A$ is the number of sites in one of the fermion reservoirs for an uncompensated mathced interface. $A_k = (2\pi/a)^2$ is the reciprocal area of the Brillouin zone of each fermion reservoir, with $a$ being the lattice constant.
We consider atomically thin films such that the Brillouin zone is 2D, and consider a quadratic dispersion of $\epsilon({\ve{k}})$ on the form 
\begin{align}
\epsilon({\ve{k}}) = \frac{\hbar^2 k^2}{2m} - \mu
\end{align}
so that the Fermi surface is a circular disk.
Here $\mu$ is the chemical potential, which we assume to be the Fermi energy $\epsilon_F$, and $m$ is the effective fermion mass.
For our simplified contact interaction (independent of momentum) the magnon potential just becomes a constant and therefore has to lead to an isotropic s-wave pairing. We can then write the momentum integral as an energy integral
\begin{align}
\int_\text{BZ}\diff^2k &\rightarrow 2\pi\int_0^{k_\text{BZ}}\diff k k \rightarrow \frac{2\pi m}{\hbar^2} \int_{-\epsilon_F}^{\epsilon_\text{BZ}-\epsilon_F} \diff \epsilon \rightarrow \frac{2\pi m}{\hbar^2} \epsilon_F \int_{-\infty}^\infty \diff x\, ,
\end{align}
where $x$ is a dimensionless energy integral, $\epsilon_F = \hbar^2k_F^2/(2m)$ is the Fermi energy, and $\epsilon_\text{BZ}=\epsilon(k_\text{BZ})$ is the energy at the Brillouin zone boundary $k=k_\text{BZ}$.
We extend the integration limits on the integral over $x$ to $\pm\infty$, as the biggest contributions come from near the Fermi surface ($x=0$), and we will just add a small error by integrating over all energies.
Noting that the energy integral can be determined to be
\begin{align}
\epsilon_F\int_{-\infty}^{\infty}\diff x \frac{1}{(\epsilon_F x)^2+\abs{\Delta_{\sigma}}^2+(\hbar\nu_n)^2} = \frac{\pi}{\sqrt{\abs{\Delta_{\sigma}}^2+(\hbar\nu_n)^2}} \, ,
\end{align}
the gap equation simplifies further to
\begin{align}
\Delta_{-\sigma}=\beta^{-1}U_{\sigma}(\ve{0},0)\Delta_{\sigma}\frac{N_A}{(2\pi/a)^2} \frac{2\pi^2 m}{\hbar^2}\sum_{n} \frac{1}{\sqrt{\abs{\Delta_{\sigma}}^2+(\hbar\nu_n)^2}} \, .
\end{align}
While the Matsubara sum in the gap equation above in principle runs over all frequencies for the approximations we have made, the sum will be divergent. To ensure that the sum converges, we introduce some upper limit to the sum. Looking at the form of the magnon-mediated potential in Eq. \eqref{eq:general_magnon_potential}, we see that the potential is only attractive in a certain range of frequencies. We assume that the interfaces are defined so that it is attractive for small frequencies, i.e. we have an interface with sublattice $A$ on one side and the other interface is with sublattice $B$. Through an analytic continuation of Eq. \eqref{eq:general_magnon_potential} one can see that the attractive frequency region is then bounded by $\omega_n = \varepsilon_{\ve{q},\mu}/\hbar$ on one side, and by $\omega_n = \varepsilon_{\ve{q},\nu}/\hbar$ on the other. As we are only looking at contact interactions ($\ve{q}=\ve{0}$), the bounds on the attractive frequency region is given by the magnon gaps $\varepsilon_{\ve{0},\mu/\nu}$. In the system we consider the magnon gaps are identical, $\varepsilon_{\ve{0},\mu}=\varepsilon_{\ve{0},\nu}\equiv\varepsilon_{\ve{0}}$. We therefore only sum over Matsubara frequencies for $n$ between $-N$ and $N$, where $N$ is defined as $\abs{\hbar\nu_{N}} = \varepsilon_{\ve{0}}$. Assuming $N$ is large, we have $N\approx\beta\varepsilon_{\ve{0}}/(2\pi)$.

At the critical temperature the gaps $\Delta_{\pm\sigma}$ vanish. We assume that the gaps have the same critical temperature $T_c$, and that they obey the limit
\begin{align}
\lim_{T\rightarrow T_c} \frac{\Delta_{-\sigma}}{\Delta_{\sigma}} = 1 \, .
\end{align}
If not exactly 1, the ratio still has to be positive for there to be a solution of the gap equation. In the ideal spin-degenerate case, it is however a sensible assumption that the ratio should be 1.
With these assumptions, the gap equation simplifies to
\begin{align}
\frac{2\hbar^2\beta}{N_A m a^2 U_{\pm\sigma}(\ve{0},0)} &\approx \sum_{n=-N}^{N}\frac{1}{\abs{\hbar\nu_n}}= \frac{\beta}{\pi}\sum_{n=0}^{N}\frac{1}{\abs{n+1/2}}
\end{align}
The sum can be approximated by \cite{kopnin2001}
\begin{align}
\sum_{n=0}^N\frac{1}{n+1/2} \approx \ln(N) +2\ln(2) + \gamma_\text{EM} \, ,
\end{align}
where $\gamma_\text{EM}=0.577\ldots$ is the Euler--Mascheroni constant.
If we consider the scenario where we have an interface with sublattice $A$ on the left side and an interface with sublattice $B$ on the right side, as this yields an attractive potential for the exciton condensation, we obtain
\begin{align}
\frac{2\pi\varepsilon_{\ve{0}}}{S u_0 v_0 m a^2 J_A^L J_B^R} \equiv \frac{1}{\lambda} \approx \ln\left(\frac{\beta\varepsilon_{\ve{0}}}{2\pi}\right) +2\ln(2) + \gamma_\text{EM} \, .
\end{align}
Exponentiating the above equation, we find the analytical expression for $T_c$:
\begin{align}\label{eq:Tc1}
T_c 
&= \frac{2e^{\gamma_\text{EM}}\varepsilon_{\ve{0}}}{\pi k_B}\exp\left(-\frac{2\pi\varepsilon_{\ve{0}}}{S u_0 v_0 m a^2 J_A^L J_B^R}\right)\, .
\end{align}
If we assume the exchange energy of the antiferromagnetic spins is much larger than the interface coupling ($\hbar\omega_E\gg S m a^2 J_A^L J_B^R$), we find that the $T_c$ is maximized for a given exchange energy by the anisotropy 
\begin{align}
\omega_\parallel^\text{(opt)} = \frac{S m a^2 J_A^L J_B^R}{16\pi\hbar} \, .
\end{align}
When the anisotropy takes on this optimal value, the critical temperature becomes
\begin{align}
T_c=\frac{\sqrt{\hbar\omega_E S m a^2 J_A^L J_B^R}}{\sqrt{2}\pi^{3/2}k_B}e^{\gamma_\text{EM}-1/2} \, .
\end{align}

\section{Retardation effects and generalizations}

In our model and analysis above, we have employed essentially the same simplifying assumptions that appear in the BCS theory~\cite{Bardeen1957} of superconductivity and obtained an analogous analytic expression for the critical temperature. The goal has been to capture the essential physics and achieve a minimalistic understanding that may guide experimental efforts towards realizing magnon-mediated exciton condensates. In the following, we discuss some of these assumptions and their validity, especially the so-called retardation effects. Furthermore, capitalizing on the extensive detailed studies of superconductivity employing the Eliashberg theory~\cite{McMillan1968,Combescot1990,Marsiglio2018}, we anticipate the corresponding corrections to our $T_c$ expression that should result from analogous investigations.

For the present discussion, it is convenient to express the critical temperature expression [Eq. (\ref{eq:Tc1})] in the standard form:
\begin{align}\label{eq:Tc2}
    k_B T_c = & 1.13 ~ \hbar \omega_c ~ \exp{ \left( - \frac{1}{\lambda} \right)  },
\end{align}
where $\lambda$ becomes the dimensionless electron-magnon coupling parameter and $\hbar \omega_c = \varepsilon_{\ve{0}}$. This BCS expression has been obtained under two mathematical assumptions: (i) the effective electron-hole attraction and the condensate order parameter are independent of Matsubara frequencies up to the cutoff $\varepsilon_{\ve{0}}/\hbar$, and (ii) only the uniform magnon modes (with energy $\varepsilon_{\ve{0}}$) contribute to the electron-hole attraction. In the absence of a cutoff, the former assumption implies that fermion-fermion interaction is instantaneous in time. Imposition of the cutoff partially accounts for the finite delay in the interaction and introduces some form of retardation effect in the analysis. The second assumption above considers that dominant contribution comes from the uniform magnon modes and thus disregards magnons with higher frequencies. This is equivalent to assuming a wavevector-independent Einstein model for the bosons~\cite{Combescot1990,Marsiglio2018}.

Employing Eliashberg theory, these assumptions can be relaxed systematically and their effects can be examined. We continue to work within the weak coupling approximation. It has been found that accounting rigorously for retardation effects, i.e. allowing fermion-fermion interaction to be frequency dependent, indeed significantly affects the frequency dependence of the order parameter~\cite{Marsiglio2018,Combescot1990}. On the other hand, the $T_c$ expression Eq. (\ref{eq:Tc2}) is altered merely via the replacement $\omega_c \to \omega_c / \sqrt{e}$ within the unrenormalized Eliashberg theory~\cite{Marsiglio2018}. Accounting for quasiparticle renomarlization effects~\cite{McMillan1968,Marsiglio2018} in addition introduces a further suppression of $T_c$ by  factor $e$ making the overall replacement $\omega_c \to \omega_c / e^{3/2}$. However, in the case of superconductivity, the electron-phonon interaction causes the electron-electron attraction as well as the quasiparticle renormalization thereby necessitating a proper inclusion of the latter in considering superconductivity~\cite{McMillan1968,Marsiglio2018}. In the case of magnon-mediated quasiparticle-quasiparticle interaction, renormalization effects may not be important for condensation and may still be dominated by the quasiparticle-phonon interaction. Therefore, for our case of magnon-mediated exciton condensation, a detailed Eliashberg theory accounting for retardation and quasiparticle renormalization effects should lead to a reduction in $T_c$ [Eq. (\ref{eq:Tc2})] by a factor between $\sqrt{e}$ and $e^{3/2}$. On the other hand, our simplifying the magnon spectrum with a single frequency at the bottom of the band underestimates the cutoff $\omega_c$ and thus $T_c$. It has been shown that the effective cutoff is obtained via log-averaging the frequency spectrum~\cite{Combescot1990}. Since our chosen cutoff of $\hbar \omega_c = \varepsilon_{\ve{0}}$ corresponds to the band bottom, this detailed averaging requires replacing the $\omega_c$ with a larger frequency, thereby increasing the $T_c$~\cite{Combescot1990}.

Thus, compared to our simplified result for $T_c$ [Eq. (\ref{eq:Tc2})], retardation and quasiparticle renormalization effects tend to reduce the $T_c$. At the same time, accounting for the full bandstructure tends to increase it. Both of these effects are expected to be comparable leaving our simple $T_c$ expression a good estimate even after including these complications~\cite{Combescot1990,Marsiglio2018}.